\documentclass[aps,twocolumn,superscriptaddress]{revtex4-2}

\usepackage{graphicx}
\usepackage{amssymb,amsmath}
\usepackage{braket}
\usepackage{placeins}

\makeatletter
\newcommand*{\addFileDependency}[1]{
\typeout{(#1)}
\@addtofilelist{#1}
\IfFileExists{#1}{}{\typeout{No file #1.}}
}\makeatother

\newcommand*{\myexternaldocument}[1]{%
\externaldocument{#1}%
\addFileDependency{#1.tex}%
\addFileDependency{#1.aux}%
}

\newcommand{\Gsqz}{G_\mathrm{S}}
\newcommand{\phiamp}{\phi_\mathrm{A}}
\newcommand{\phisqz}{\phi_\mathrm{S}}
\newcommand{\Idc}{I_\mathrm{DC}}
\newcommand{\PN}{P_\mathrm{N}}
\newcommand{\Psa}{P_{\mathrm{N}}^{\mathrm{S,A}}}

\newcommand{\Pa}{P_{\mathrm{N}}^{\mathrm{A}}}

\newcommand{\Th}{T_\mathrm{H}} 
\newcommand{\Tmc}{T_\mathrm{MC}} 
\newcommand{\varsa}{\delta I^2_\mathrm{S,A}} 
\newcommand{\vara}{\delta I^2_\mathrm{A}} 
\newcommand{\Sth}{S_\mathrm{th}} 

\usepackage{xr}
\myexternaldocument{supplement}

\usepackage{xcolor}
\usepackage{soul}

\usepackage{hyperref}

\begin{document}


\title{Strong Microwave Squeezing Above 1 Tesla and 1 Kelvin}

\author{Arjen Vaartjes}
\thanks{These authors contributed equally}
\affiliation{School of Electrical Engineering and Telecommunications, UNSW Sydney, Sydney, NSW 2052, Australia}
\author{Anders Kringh\o{}j}
\thanks{These authors contributed equally}
\affiliation{School of Electrical Engineering and Telecommunications, UNSW Sydney, Sydney, NSW 2052, Australia}
\author{Wyatt Vine}
\thanks{These authors contributed equally}
\affiliation{School of Electrical Engineering and Telecommunications, UNSW Sydney, Sydney, NSW 2052, Australia}
\author{Tom Day}
\affiliation{School of Electrical Engineering and Telecommunications, UNSW Sydney, Sydney, NSW 2052, Australia}
\author{Andrea Morello}
\affiliation{School of Electrical Engineering and Telecommunications, UNSW Sydney, Sydney, NSW 2052, Australia}
\author{Jarryd J. Pla}
\email[]{jarryd@unsw.edu.au}
\affiliation{School of Electrical Engineering and Telecommunications, UNSW Sydney, Sydney, NSW 2052, Australia}
\date{\today}

\begin{abstract}
Squeezed states of light have been used extensively to increase the precision of measurements, from the detection of gravitational waves~\cite{Aasi2013} to the search for dark matter~\cite{Malnou2019, Backes2021}. In the optical domain, high levels of vacuum noise squeezing are possible due to the availability of low loss optical components and high-performance squeezers. At microwave frequencies, however, limitations of the squeezing devices~\cite{Boutin2017, Bienfait2017, Malnou2018} and the high insertion loss of microwave components~\cite{Malnou2019} makes squeezing vacuum noise an exceptionally difficult task. Here we demonstrate a new record for the direct measurement of microwave squeezing. We use an ultra low loss setup and weakly-nonlinear kinetic inductance parametric amplifiers~\cite{Parker2021} to squeeze microwave noise 7.8(2)~dB below the vacuum level. The amplifiers exhibit a resilience to magnetic fields and permit the demonstration of record squeezing levels inside fields of up to 2~T. Finally, we exploit the high critical temperature of our amplifiers to squeeze a warm thermal environment, achieving vacuum level noise at a temperature of 1.8~K. These results enable experiments that combine squeezing with magnetic fields~\cite{Malnou2019, Bienfait2017, Vine2023} and permit quantum-limited microwave measurements at elevated temperatures, significantly reducing the complexity and cost of the cryogenic systems required for such experiments.
\end{abstract}

\maketitle

\section{Introduction}
The measurement of weak signals is at the heart of many important challenges in modern science and engineering, from quantum computing~\cite{Gambetta2017}, to spectroscopy~\cite{Bienfait2016}, to the search for gravitational waves and dark matter~\cite{Aasi2013, Backes2021}. Ultimately, the ability to measure a weak signal is constrained by noise, which at the quantum limit is dictated by vacuum fluctuations of the electromagnetic field. Vacuum fluctuations are a manifestation of the uncertainty principle, where the two quadrature components of a signal $I$ and $Q$ must obey the relation $\delta I^2\delta Q^2 > 1/16$, where $\delta I^2$ ($\delta Q^2$) is the variance of the signal in the $I$ ($Q$) quadrature, measured in the unitless dimension of photons. This simple inequality provides a way to circumvent the seemingly fundamental constraint imposed by vacuum fluctuations: the noise along one quadrature can be reduced beneath the quantum limit, so long as the noise is increased in the opposite quadrature. This process is known as squeezing and can be exploited to enhance the signal-to-noise ratio (SNR) of a measurement. 

A common approach to produce a squeezed state is to use a degenerate parametric amplifier (DPA) to amplify a vacuum state~\cite{Malnou2018}. A DPA exhibits phase-sensitive gain, such that it amplifies noise along one quadrature and deamplifies (or squeezes) it along the other~\cite{Castellanos2008, Bienfait2017, Malnou2019, Parker2021}. It is essential that the DPA behave as a noiseless (i.e. ideal) amplifier, as any noise added will degrade the squeezed state. Moreover, loss between the DPA and the detection apparatus also diminishes squeezing and must be minimized. These requirements make producing highly squeezed states an exceptionally difficult task. 

In the optical domain, the availability of high-performance DPAs, ultra-low loss optical components and the relative ease of noiseless optical homodyne measurements have facilitated the demonstration of vacuum states squeezed by as much as 15~dB~\cite{Vahlbruch2016}. At microwave frequencies, Josephson junction-based parametric amplifiers represent the state-of-the art in microwave noise squeezers~\cite{Backes2021, Qiu2023}. However, higher order nonlinearities present in Josephson parametric amplifiers (JPAs) constrains the amount of squeezing that can be achieved~\cite{Boutin2017, Liu2017, Bienfait2017, Malnou2018, Frattini2018}. In addition, microwave components are lossy in comparison to their optical counterparts and microwave signals at the single photon level require near-noiseless amplification prior to measurement with conventional electronics~\cite{Eichler2012}. As a result, many works have reported only an inferred measurement of squeezing~\cite{Castellanos2008, Zhong2013, Qiu2023}, where degradation of the squeezed state between the DPA and detection system is not accounted for. Due to the additional complexity, there are few reports of direct measurements of microwave noise squeezing, where notably 4.5~dB of squeezing was utilized in an axion haloscope to speed up the search for dark matter~\cite{Malnou2019}.

\begin{figure*}[t!]
\includegraphics[width=189mm]{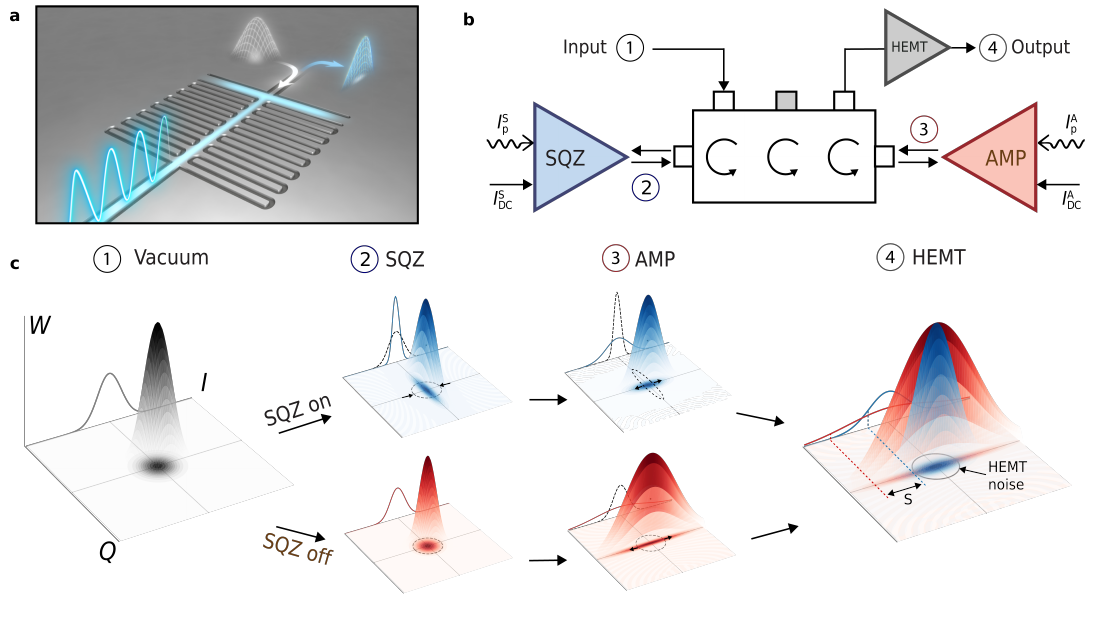} 
\caption{\textbf{Setup for a direct measurement of microwave vacuum squeezing.} \textbf{a}, Artist's impression of the squeezed state generation inside a two-port kinetic inductance parametric amplifier (KIPA). On the left, an AC pump tone (blue wave) and a DC current (blue glow) enter a half-wavelength coplanar waveguide (CPW) cavity with interdigitated capacitance. The DC current is shunted to ground via an inductance (blue glow). On the right side, a vacuum state (gray) gets reflected from the KIPA as a squeezed state (blue). \textbf{b}, Two KIPAs are connected via a triple-junction circulator in a squeezed state receiver setup, with one labeled squeezer (SQZ) and the other labeled amplifier (AMP). The KIPAs are controlled with independent DC $\Idc$ and pump $I_\mathrm{p}$ signals. \textbf{c}, The procedure used to achieve a direct measurement of the squeezing produced by SQZ consists of two independent measurements. (1) A vacuum state is supplied to the input of the three-junction circulator and reflected off SQZ. (2) When SQZ is on (blue, top row) the state is squeezed along the $I$-quadrature. When SQZ is off (red, bottom row), it remains a vacuum state. (3) AMP is used to anti-squeeze (amplify) the resulting state along $I$, before the state is amplified by a high electron mobility transistor (HEMT) amplifier (4). The level of vacuum squeezing $S$ is defined by the difference (in decibels) of the noise power measured along $I$ at room temperature in the two scenarios (SQZ on and SQZ off). The 3D-plots represent the Wigner function $W$ of the various states.}
\label{fig1}
\end{figure*}

Here we use a new microwave squeezer made from a thin film of niobium titanium nitride (NbTiN), a material that exhibits a weak nonlinearity due to its kinetic inductance~\cite{Parker2021}. Unlike previous Josephson junction-based microwave squeezers, the kinetic inductance parametric amplifier (KIPA) is compatible with high magnetic fields and high temperatures and has negligible higher order nonlinearities~\cite{Parker2021}. We combine the KIPA with commercial microwave components and custom engineered device enclosures to minimize microwave losses, allowing us to achieve a direct measurement of $-7.8(2)$~dB of microwave vacuum noise squeezing, the highest amount reported to date. In addition, we show that high levels of squeezing can be maintained with the device operated in an in-plane magnetic field of up to 2~T, which permits its integration in applications such as spin resonance spectroscopy~\cite{Bienfait2017, Vine2023} and axion detection~\cite{Malnou2019, Backes2021}. Finally, the high critical temperature ($\sim 13$~K) of the superconducting film allows the squeezing of thermal states with the device operated at elevated temperatures, enabling vacuum level noise at a frequency of 6.2~GHz and a temperature of 1.8~K. These results show that quantum-noise-limited measurements can be performed at temperatures accessible using affordable pumped Helium-4 cryostats, a noise level that would otherwise only be attainable with comparatively expensive dilution refrigerator technology.

\begin{figure}[t!]
\includegraphics{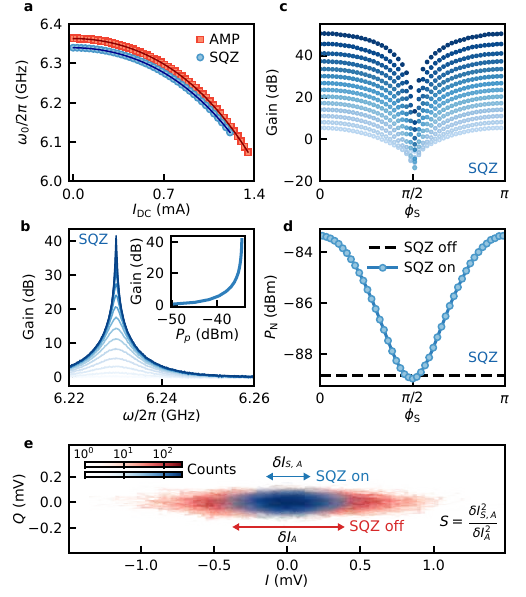}
\caption{\textbf{Amplification and Squeezing with KIPAs.} \textbf{a}, DC current $\Idc$ dependence of the resonance frequency $\omega_0$ of SQZ (blue) and AMP (red). Solid lines are fits to theory (see Supplementary Eq.~S38). \textbf{b}, Phase-insensitive gain of SQZ as a function of probe frequency $\omega/2\pi$ with a pump tone applied at $\omega_p/2\pi=2\omega_0/2\pi=12.46$~GHz and $\Idc=0.90$~mA. Increasing opacity corresponds to increasing the pump power from $P_p=-49$ to $-34$~dBm. Inset: Extracted maximum gain as a function of $P_\mathrm{p}$. \textbf{c}, Phase-sensitive gain at $\omega/2\pi=6.23$~GHz with $\omega_p=2\omega_0$ for varying pump phase $\phisqz$. Increasing opacity corresponds to increasing the pump power from $P_p=-49$ to $-33$~dBm. \textbf{d}, Comparison of noise power $\PN$ for an input vacuum state with SQZ on (blue data points) compared to SQZ off (black dashed line) as a function of $\phisqz$. For $\phisqz = \pi/2$, maximum deamplification occurs, resulting in a noise level $0.14$~dB below the reference system noise, which is measured with SQZ turned off. AMP is off for both measurements. \textbf{e}, Histograms of the $I$ and $Q$ quadrature signals measured with only AMP on (red data points) and with both SQZ and AMP on (blue data points). The blue and red arrows show the standard deviations of the distributions along the $I$-quadrature: $\varsa$ and $\vara$, respectively. The ratio of the variances corresponds to a direct measure of the achieved squeezing: $S=\varsa/\vara$.
}
\label{fig2}
\end{figure}

\section{Squeezing setup}
The KIPA devices exploit a weakly nonlinear kinetic inductance, which enables a three-wave mixing (3WM) interaction~\cite{Parker2021} in a half-wavelength coplanar waveguide (CPW) resonator (see Fig.~\ref{fig1}a). Each KIPA is operated by applying a DC current $\Idc$ and a pump tone $I_p\sin(\omega_pt+2\phi_p)$ at approximately twice the device's resonance frequency $\omega_p\approx2\omega_0$. The phase of the pump tone $\phi_p$ determines along which axis the KIPA squeezes. The DC current serves two purposes: it tunes the resonance frequency $\omega_0$ of the KIPA~\cite{Annunziata2010,Vissers2015}, and it strengthens the 3WM interaction, whilst leaving the Kerr strength negligibly low~\cite{Vissers2016}. As a result, we strongly suppress higher order nonlinearities that have previously limited microwave squeezing experiments utilizing JPAs~\cite{Malnou2018, Boutin2017}. 

To facilitate direct measurements of microwave squeezing, we employ an experimental setup known as a squeezed state receiver (SSR)~\cite{Malnou2019}, with special care taken to minimize insertion loss and noise in the setup. In our experiments, the SSR is composed of two KIPAs with only a triple-junction circulator between them (Fig.~\ref{fig1}b). To enable this configuration, we modify the KIPA design from those used in previous works~\cite{Parker2021,Vine2023} by adding a second port, allowing the pump and signal tones to be routed independently. We also make use of on-chip (superconducting) filters to suppress pump leakage (see Supplementary Information Section III.B for further details). The two KIPAs are nominally identical but serve two distinct purposes in our experiments: we label the first SQZ because it is used to produce squeezed states from vacuum; we label the second AMP because it is used to noiselessly amplify the squeezed states. Both devices are installed in a dilution refrigerator with a base temperature of 10~mK. 

We measure squeezing with a two-step protocol, displayed in Fig.~\ref{fig1}c. In the first step SQZ and AMP are both activated, but are made to amplify along orthogonal quadratures, i.e., SQZ produces a squeezed vacuum state aligned on the $I$-quadrature, whereas AMP anti-squeezes along $I$. We then measure the variance on $I$, $\varsa$. In the second step, we deactivate SQZ and make a measurement of the variance on same quadrature $\vara$. Since the only difference between the two steps is the squeezing of the vacuum state, the vacuum squeezing level can be found from the ratio of the two variances $S=\varsa/\vara$.

\section{Operation of the parametric amplifiers}
Essential to the measurement protocol is that SQZ and AMP operate at the same frequency. This can be achieved over a 215~MHz span by using DC currents $\Idc$ to tune the two devices to a mutual resonance frequency $\omega_0$ (Fig.~\ref{fig2}a). At $\omega_0/2\pi=6.23~$GHz we measure the phase-insensitive gain of each KIPA by supplying a pump tone with frequency $\omega_p=2\omega_0$ and varying the frequency $\omega$ of a probe tone. In Fig.~\ref{fig2}b we show the measurement for SQZ where a maximum power gain of 41.5~dB was achieved for a pump power of $P_{\mathrm{p}}^\mathrm{S}=-33~$dBm, with a gain bandwidth product of $2\pi\times 17~$MHz. For phase-sensitive amplification, relevant for squeezing, we amplify coherent states with frequency $\omega_0$ while keeping the frequency of the pump tone fixed at $\omega_p=2\omega_0$. Figure~\ref{fig2}c shows that the gain of SQZ varies between $50~$dB and $-14$~dB as a function of the pump phase $\phisqz$. The performance of AMP is found to be nearly identical to SQZ (see Supplementary Information Section IV.C).

In the absence of a probe signal, the field input to SQZ is a vacuum state. By measuring the phase-dependent noise power $\PN$ along the $I$-quadrature, which is directly proportional to the variance of the noise, we demonstrate the squeezing capability of SQZ, with AMP deactivated. When $\phisqz \approx \pi/2$, the noise level (blue curve in Fig.~\ref{fig2}d) falls slightly below the reference system noise, which is measured when both SQZ and AMP are deactivated (black dashed line in Fig.~\ref{fig2}d). Notably, the amount of squeezing observed is modest (-0.14~dB). This is because the next activated amplifier in the detection chain -- a high electron mobility transistor (HEMT) amplifier -- adds 6.9 photons of noise to each quadrature (see Supplementary Information Section IV.G), overwhelming the squeezed noise. Even if the following amplifier were quantum-limited but operated in non-degenerate mode (where a minimum of 0.25 photons of noise is added to each quadrature~\cite{Caves1982}), the maximum amount of squeezing that could be achieved is -3~dB. This emphasizes the need for a second DPA in direct measurements of vacuum squeezing, because it can boost the power of the squeezed state above the noise added by the HEMT without introducing additional noise.

Figure~\ref{fig2}f presents a direct measurement of squeezing utilizing both SQZ and AMP, as outlined in the protocol described above. The measured noise is plotted as histograms in the $IQ$-plane, constructed from time traces of the noise 
(see Methods). The pump phases of SQZ and AMP are set to $\phi_\mathrm{S}=\pi/2$ and $\phi_\mathrm{A}=0$ so that they maximally squeeze and amplify along $I$, respectively. When SQZ is deactivated (Fig.~\ref{fig2}e, red data points), the vacuum state is amplified by AMP, resulting in an ellipse whose major axis is aligned along $I$. When SQZ is activated (Fig.~\ref{fig2}e, blue data points), the variance along $I$ shrinks by an amount corresponding to the achieved vacuum squeezing level, $S$.

\section{Optimization of Direct Squeezing}

\begin{figure}[t!]
\includegraphics{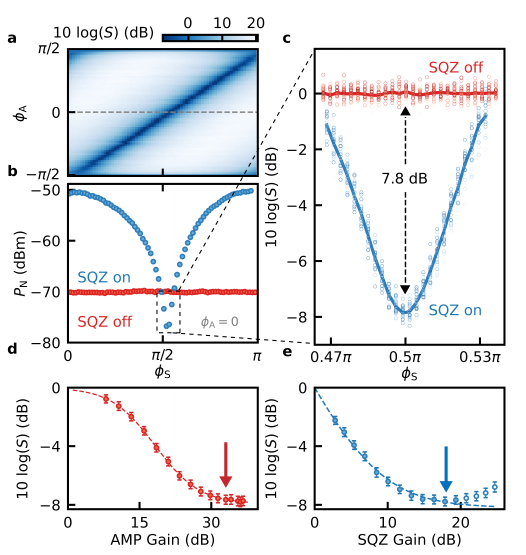}
\caption{\textbf{Squeezing optimization.} \textbf{a}, Squeezing $S$ as a function of pump phases $\phisqz$ and $\phiamp$. The diagonal feature corresponds to SQZ and AMP squeezing along orthogonal axes, where $S$ is optimal. \textbf{b}, A linecut at $\phiamp = 0$ in panel a, showing the noise power $P_{N}$ on the $I$-quadrature in two scenarios: `SQZ on' and `SQZ off' (AMP is always on). Around $\phisqz = \pi/2$, SQZ reduces the noise on $I$ below the reference vacuum level. \textbf{c}, A zoom-in around $\phisqz = \pi/2$, where SQZ is maximally squeezing along the $I$-quadrature. The squeezed noise reaches a minimum of $7.8(2)$~dB below the vacuum level. \textbf{d, e}, Maximum squeezing as a function of AMP and SQZ gain. The gain of both amplifiers was measured via the anti-squeezing of a coherent state. Arrows indicate the setpoints used for the other panels. The dashed lines are fits to a cavity input-output model, which is derived in Supplementary Information Section II.B. The un-filled data points in \textbf{e} deviate from the model and are excluded from the fit.}
\label{fig3}
\end{figure}



To determine the maximum achievable squeezing in our system, we require $\phisqz$ and $\phiamp$ to be aligned precisely, such that the two devices amplify along orthogonal axes. As expected, we observe high levels of squeezing whenever $\phisqz \approx \phiamp + \pi/2$ (Fig.~\ref{fig3}a). At the optimal setpoint, we observe a maximum squeezing of $10\log(S)=-7.8(2)$~dB (Figs.~\ref{fig3}b,c). We emphasize that this level of squeezing is not an inferred value, but rather a direct measurement of an itinerant squeezed state without correcting for loss or system noise. This amount of squeezing is the highest value recorded to date in a direct measurement at microwave frequencies. 


Squeezing is found to saturate with the gain of both amplifiers (Figs.~\ref{fig3}d,e). This can be explained using a model based on input-output theory which captures several distinct mechanisms that limit $S$ (see Supplementary Information Section II.B). In the high AMP gain limit, the squeezing level becomes

\begin{equation}
    S = 1 - \frac{\eta_1(1-\Gsqz)(1/4-n^\mathrm{sq}_{\mathrm{S}})}{1/4+n^\mathrm{anti}_{\mathrm{A}}},
    \label{eq:sqz}
\end{equation}

\noindent where $n^\mathrm{sq}_{\textrm{S}}$ is the number of noise photons added by SQZ to $I$ when it is squeezing, $n^\mathrm{anti}_{\textrm{A}}$ is the number of noise photons added by AMP to $I$ when it is anti-squeezing, and $\eta_1$ is the transmission efficiency. Fitting our measurements to the full model (Supplementary Eq.~S21) yields a combined factor $\eta_1(1/4 - n^\mathrm{sq}_{S})/(1/4 + n^\mathrm{anti}_A) = 0.85$. From independent measurements of the amplifier added noise we determine $n^\mathrm{sq}_{\textrm{S}}=0.02(2)$ and $n^\mathrm{anti}_{\textrm{A}}=0.00(2)$ (see Supplementary Information Section IV.G), which yields $\eta_1=0.92(8)$, equivalent to $-0.34$~dB of insertion loss. This finding is in close agreement with our estimates based on the components in our setup, as detailed in the Supplementary Information (Section I.B).

What is not captured by the model is the degradation of $S$ observed when the gain of SQZ exceeds $20~$dB, which may be explained by an increased sensitivity of the deamplification gain to small drifts in $\phisqz$ as the pump power is increased (see Fig.~\ref{fig2}c and Supplementary Fig. S13). 

\section{Squeezing in Magnetic Fields and at Elevated Temperatures}

\begin{figure*}[t]
\includegraphics[width=\textwidth]{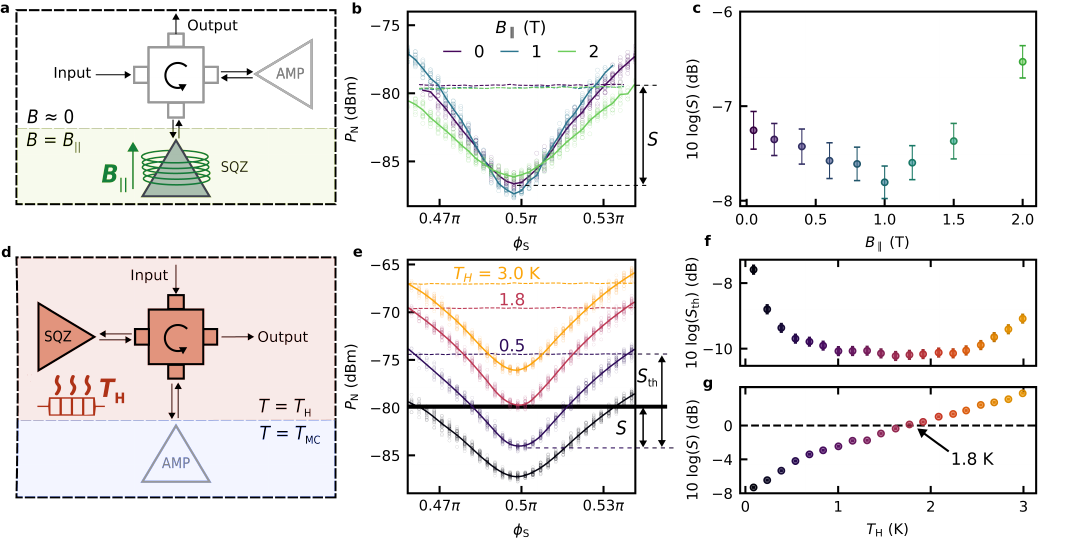}
\caption{\textbf{Magnetic field resilience and thermal squeezing.} \textbf{a}, Modified squeezing setup, where SQZ is placed in an in-plane magnetic field $B_{\parallel}$. All other components are magnetically shielded such that $B\approx0$~T. \textbf{b}, Noise reduction with respect to the reference vacuum level (dashed lines) for fields up to $B_{\parallel} = 2$~T. \textbf{c}, Vacuum squeezing as a function of magnetic field $B_{\parallel}$. The squeezing level remains below $-6.3$~dB up to 2~T. \textbf{d}, Modified setup for thermal squeezing. The circulator and SQZ are thermally anchored to a heat source with variable temperature $\Th$. AMP is thermally isolated at $T=\Tmc \approx 10$~mK. \textbf{e}, Noise squeezing for several values of $\Th$. The horizontal dashed lines indicate a rise in noise power due to the elevated temperature. The second line from the top (pink) shows that thermal noise at $\Th=1.8$~K is squeezed down to the vacuum level (black horizontal solid line). The vertical arrows indicate the definitions of vacuum squeezing $S$ and thermal squeezing $\Sth$. \textbf{f}, Squeezing $\Sth$ with respect to a thermal input state. Note that the thermal squeezing level reaches $-10$~dB; in this configuration the squeezing is primarily limited by the insertion loss between SQZ and AMP. Above $\Th \approx 2.25$~K, $\Sth$ begins to degrade due to a reduction in the internal quality factor of SQZ (see Supplementary Information Section IV.F). \textbf{g}, Squeezing with respect to the vacuum $S$ as a function of heater temperature $\Th$. At $\Th = 1.8$~K, the squeezed noise power is equal to the power of the vacuum fluctuations. The error bars in \textbf{g} are smaller than the markers.}
\label{fig5}
\end{figure*}

\textbf{Magnetic Field} Several applications requiring strong magnetic fields benefit from the use of squeezed states, including electron spin resonance spectroscopy~\cite{Bienfait2017, Eichler2017} and dark matter axion detection~\cite{Backes2021}. Previous works have utilized JPAs but have required the amplifiers to be magnetically shielded. In contrast, several recent works have demonstrated magnetic field-resilient resonant amplifiers based on materials with kinetic inductance~\cite{Xu2023, Splitthoff2023}, though it remains an outstanding goal to realize highly squeezed states in such an environment.

To demonstrate the magnetic field-resilience of our squeezing setup, we modify our experiment so that SQZ is exposed to a magnetic field $B_{\parallel}$, while the rest of the components (including AMP and the circulator) are magnetically shielded (Fig.~\ref{fig5}a). $B_\parallel$ is aligned to be in-plane with SQZ, to minimize the creation of magnetic flux vortices, which is expected to introduce additional loss in the resonator~\cite{Bothner2012}. Figure~\ref{fig5}b and Fig.~\ref{fig5}c show that the squeezing level remains below $-7$~dB up to $B_{\parallel}=1.5$~T. This level of squeezing is slightly less than was demonstrated in Fig.~\ref{fig3}c, which might be explained by a slight shift of the optimal operation point in the modified setup. 

The degree of squeezing we achieve is initially found to increase with $B_\parallel$, which is paired with an increase in $Q_i$ (see Supplementary Fig.~S10). A similar effect was observed in Ref.~\cite{Kroll2019} for NbTiN resonators measured in in-plane magnetic fields. As $B_\parallel$ is raised beyond 1~T, however, $S$ becomes limited by the internal quality factors of SQZ and AMP, both of which begin to degrade. Interestingly, the maximum field at which we measured squeezing (2~T) is limited by AMP, not SQZ. At larger fields, AMP loses superconductivity as a result of the stray magnetic field, which is a factor of $\sim 200$ less than $B_\parallel$, but is not aligned in-plane. This implies that a high degree of squeezing can be achieved at fields greater than 2~T with careful alignment of the magnetic field with both amplifiers. \\

\textbf{Thermal squeezing} Next, we leverage the high critical temperature of our amplifiers to demonstrate squeezing of the noise produced by a hot thermal bath down to the vacuum level. By heating up not only the input state, but also SQZ and the circulator, we demonstrate that our setup is capable of reducing thermal microwave noise to the quantum limit, at temperatures up to 1.8~K.

We detect the squeezed thermal states with AMP, which is mounted on the mixing chamber of the dilution refrigerator and remains at its base temperature $\Tmc\approx 10~$mK. In contrast, SQZ and the circulator are both thermally anchored to a variable thermal source (VTS) which is heated to a temperature $\Th$ (Fig.~\ref{fig5}d). Raising $\Th$ generates a thermal state containing $n$ photons that is directed to the input of SQZ. The total photon number is given by $n = n_{\mathrm{th}} + 1/2$, where $n_{\mathrm{th}} = 1/[\exp(\hbar\omega/k_\mathrm{B} \Th) - 1]$ is the Bose-Einstein distribution, with $k_{\mathrm{B}}$ the Boltzmann constant and $\hbar$ the reduced Planck constant.

In Fig.~\ref{fig5}e, we present measurements performed at four values of $\Th$. At each setpoint, we measure the noise power of the signal with both amplifiers activated ($\Psa$) and when only AMP is activated ($\Pa$). The latter is indicated in Fig.~\ref{fig5}e as a series of dashed lines, which rise in power with $\Th$ because of the increased thermal noise $n(\Th)$. At each temperature, $\Psa$ can be reduced below $\Pa$ for pump phases $\phisqz \approx \pi/2$. We define this reduction to be the level of thermal squeezing, $\Sth$. In Fig.~\ref{fig5}f we show that $\Sth$ initially increases with $\Th$, and saturates at $10\log(\Sth) = -10.2(1)$~dB for $\Th$ in the range $1-2~$K. As $\Th$ is increased further, $\Sth$ begins to deteriorate, which we attribute to a decline in the internal quality factor $Q_\mathrm{i}$ of SQZ with $\Th$ (see Supplementary Fig.~S11). 

We note that the maximum level of thermal squeezing achieved here is greater than the vacuum squeezing shown in Fig.~\ref{fig3}c. This is because as the temperature increases, the impact of KIPA-added noise on squeezing reduces (see Supplementary Information Section II.D), in which case the squeezing is predominantly limited by the insertion loss between SQZ and AMP. This allows us to estimate the transmission efficiency between SQZ and AMP as $\eta_1 = 0.927(3)$ for this experiment, which translates to an insertion loss of $-0.33$~dB (see Supplementary Fig. S14). Although the experimental setup depicted in Fig.~\ref{fig5}d differs from the one in Fig.~\ref{fig1}b in that it includes an additional superconducting coaxial cable between SQZ and AMP, the extracted insertion loss is in close agreement with the earlier found value of $-0.34$~dB.

The squeezed thermal states can also be measured relative to the vacuum level (black horizontal line in Fig.~\ref{fig5}e). By comparing the measurements of $\Psa$ taken at temperature $\Th$ with the vacuum reference (see 
Supplementary Fig. S15), we show that $S$ degrades monotonically with $\Th$ (Fig.~\ref{fig5}g). At $\Th=1.8~$K, the level of squeezing is $10\log(S)=0$~dB, indicating that the noise of the thermal state has been squeezed to the vacuum level.

\section{Discussion}
There are several avenues available for improving the amount of vacuum squeezing achieved. A significant limitation is the insertion loss of the microwave components that the itinerant squeezed state traverses~\cite{Malnou2019}. The data in Fig.~\ref{fig5}g suggests up to -10~dB of squeezing could be reached with the existing setup if the insertion loss were the only factor degrading squeezing. Whilst already low, the $-0.34$~dB of insertion loss could be reduced further by integrating the squeezing devices within the same enclosure as the circulator, circumventing the need for connectors along the squeezed state path, or fully integrating the two KIPAs using on-chip circulators~\cite{Chapman2017, Navarathna2023}. 

A second factor that limits vacuum squeezing is the noise added by the two KIPAs. Given the measured noise performance of SQZ and AMP, squeezing is limited to $-11.0$~dB in the absence of insertion loss (Eq.~\ref{eq:sqz}). This indicates that insertion loss and added noise combine almost equally to limit squeezing in our experiment. The KIPA added noise could be reduced further by carefully impedance-matching the pump port and the cavity or by reducing the impedance of the cavity, thereby lowering the required pump power and any associated heating. 

Phase instability in the setup can also limit squeezing. Any deviation from the optimal squeezing angle $\phisqz = \pi/2$ will mix in gain-dependent anti-squeezed noise, with this penalty becoming increasingly severe as the squeezer gain is increased. This is a known source of squeezing degradation in optical setups and techniques that improve phase stability (e.g. phase-locked loops) could also be employed in future microwave squeezing experiments~\cite{Lough2021}.

The high levels of microwave squeezing already demonstrated here inside magnetic fields and at high temperatures could find immediate application in electron spin resonance (ESR) spectroscopy \cite{Bienfait2017, Vine2023}, permitting quantum-limited spin detection at temperatures consistent with conventional ESR spectroscopy operating conditions ($\gtrsim 2$~K), but without the need for expensive cryogenic systems. Similarly, the search for dark matter axions \cite{Malnou2019, Backes2021} could be sped up by a factor of six compared to the quantum limit, or axion detectors could be simplified (negating the need for extensive magnetic field compensation and shielding) and made more widely available by installing them in low-tech pumped Helium-4 cryostats.

Squeezed vacuum states are also a valuable resource in measurement-based computation schemes using entangled cluster states encoded in the modes of an electromagnetic field \cite{Menicucci2006, Asavanant2019}. Critically, it has been shown that fault-tolerance with this approach can be attained using vacuum states squeezed by at least 10 dB \cite{Tzitrin2021}, which is within reach of our current setup. Superconducting circuits also offer the ability to deterministically prepare non-Gaussian states of light \cite{Campagne2020}, an essential resource in such schemes.

\section{Conclusion}
We demonstrated a record level of -7.8(2) dB of squeezing in the microwave regime using kinetic inductance parametric amplifiers. Besides the ease of fabrication, the single NbTiN layer design of these devices make the squeezers uniquely resilient against magnetic fields to at least $B_{\parallel}=2$~T. Additionally, we can employ the temperature resilience of our squeezers to achieve a noise level equivalent to vacuum with the device operated 1.8~K. This design of microwave squeezer enables quantum limited measurements at elevated temperatures, offering an alternate pathway for detecting ultra-weak microwave signals to brute-force cooling with dilution refrigerators.


\section{Methods}

\textbf{KIPA Hamiltonian}
The Hamiltonian governing the operation of the KIPA is given by

\begin{equation}
    H_{\mathrm{KIPA}}/\hbar = \Delta \hat{a}^\dagger\hat{a} + \underbrace{\frac{\xi}{2}\hat{a}^{\dagger 2}+ \frac{\xi^*}{2}\hat{a}^2 }_{\text{3WM}}+ \underbrace{\frac{K}{2} \hat{a}^{\dagger 2}\hat{a}^2}_{\text{4WM}},
    \label{eq:kipa_hamiltonian}
\end{equation}

\noindent which is defined in a frame rotating at half of the pump frequency $\omega_p/2$~\cite{Parker2021}. Here $\Hat{a}$ ($\Hat{a}^{\dagger}$) is the bosonic annihilation (creation) operator, $\Delta=\omega_0-\omega_p/2$ is the frequency detuning, $\xi = -e^{-i2\phi_p}\omega_0\Idc I_p/(4I_*^2)$ is the 3WM strength, where $I_*$ determines the
susceptibility to a DC current. $K$ is the self-Kerr interaction responsible for four wave mixing (4WM). The ratio $|\xi/K|$, which is used to indicate how susceptible a 3WM device is to higher-order nonlinearities~\cite{Boutin2017}, can be as large as $10^8$~\cite{Parker2021}.

\textbf{Device fabrication}
The devices are fabricated on 15~nm thick films of NbTiN deposited on an intrinsic, high-resistivity silicon chip. The devices are patterned using electron beam lithography followed by a reactive ion etch with an CF\textsubscript{4}:Ar plasma. The devices both consist of four main components: the first is a band-stop stepped-impedance filter constructed from a series of nine quarter-wavelength ($\lambda/4$) CPW segments with alternating high- ($Z_\mathrm{hi}$) and low-impedance ($Z_\mathrm{lo}$), through which we introduce the DC current $\Idc$ and pump $I_p$. The second component is a half-wavelength ($\lambda/2$) resonator, based on an interdigitated capacitor design. The third component is a fifth-order low-pass Chebyshev stepped impedance filter, which is constructed from five alternating $Z_\mathrm{hi}$ and $Z_\mathrm{lo}$ segments of varying lengths, through which we introduce the signal. The on-chip low-pass filter is crucial to avoid KIPA pump tone cross-talk. The final component is a shunt inductance to ground, which dictates the coupling quality factor $Q_c$ of the resonator. The inductance is chosen so that $Q_c\approx 200$, thereby ensuring the resonator is strongly over-coupled to the signal port. See Supplementary Information Section III.B for further details on the device design and simulations. 

\textbf{Device packaging} The two KIPA devices were mounted to their own enclosures. Both enclosures were constructed from gold-plated oxygen-free copper and act as a rectangular waveguide whose cutoff frequency is greater than the KIPAs, thereby suppressing radiation losses. The two ports of each KIPA were wire bonded to minimally-sized printed circuit boards, to minimize dielectric and ohmic losses. The devices were connected in series via a low-loss triple junction circulator. All measurements were completed with the devices mounted to the mixing chamber of a dilution refrigerator with a base temperature of 10~mK. See Supplementary Information Section I for further details on the device enclosure, mounting, and fridge wiring. 

\textbf{Homodyne reflection measurements}
Measurements of the devices are performed in reflection and recorded with a vector network analyzer (VNA), a spectrum analyzer or an oscilloscope.

We use a VNA to measure the resonance frequency and quality factors of the resonators, which we obtain from fits of $S_{11}(\omega)$ to a cavity input-output model (see Supplementary Information Section IV.A). The phase-insensitive gain of both SQZ and AMP are acquired with the same setup.

We acquire the noise ellipses in Fig.~\ref{fig2}e by demodulating the output noise with an $IQ$-mixer and recording the $I$ and $Q$ time traces on an oscilloscope with a sampling rate of 8.33~MSa/s. We apply a digital low-pass filter with a cutoff frequency of 150~kHz (the same cutoff frequency as in the squeezing measurements of Figs.~\ref{fig3}-\ref{fig5}) to filter out noise outside of the bandwidth of SQZ and AMP. We then bin the time traces to create a histogram in the $IQ$-plane.

The noise squeezing measurements are performed with a spectrum analyzer by measuring the noise power $\PN$ of one quadrature of the homodyne-demodulated output signal. We measure  the demodulated noise on $I$ with a bandwidth resolution of 300~Hz and average over a bandwidth of 120~kHz (30~kHz-150~kHz). An empirical definition of squeezing is $S=\Psa/\Pa$ where $\Psa$ and $\Pa$ are the noise powers measured when both SQZ and AMP are activated and when only AMP is activated, respectively.

We find the maximum squeezing level by applying a simple protocol. First, we set $\phiamp = 0$ to maximize $\Pa$. We then turn on SQZ and measure $\Psa(\phisqz)$ to estimate the $\phisqz$ that minimizes $S$. We take repeated measurements across a small range of $\phisqz$ and record $S(\phisqz)$ using interleaved measurements of $\Pa$ and $\Psa$. We account for phase drifts in the microwave pump sources by an alignment procedure displayed in Supplementary Fig. S13. 

\textbf{Measurements with magnetic fields and elevated temperatures} For the experiments where SQZ is measured in a magnetic field, a 6-1-1 vector magnet is installed in the dilution refrigerator. SQZ is placed in the center of the magnet solenoid, on the cold finger of the dilution refrigerator and connected to the circulator via a short length of semi-rigid NbTi superconducting cable, with the setup otherwise identical compared to the experiments performed at zero field. AMP and the circulator, situated on the mixing chamber plate, are partially magnetically shielded by a lead enclosure. 

For the experiments where SQZ is used to squeeze thermal states, SQZ and the circulator are thermally anchored to the thermal noise source, while AMP is thermally anchored to the mixing chamber plate and connected to the circulator via a a short length of semi-rigid NbTi superconducting cable.

\section{Acknowledgments}
J.J.P. acknowledges support from an Australian Research Council Discovery Early Career Research Award (DE190101397). J.J.P. and A.M. acknowledge support from the Australian Research Council Discovery Program (DP210103769). A.M. is supported by the Australian Department of Industry, Innovation and Science (Grant No. AUS-MURI000002). A.K. acknowledges support from the Carlsberg Foundation. A.V., W.V. and T.D. acknowledge financial support from Sydney Quantum Academy, Sydney, NSW, Australia. This research has been supported by an Australian Government Research Training Program (RTP) Scholarship. The authors acknowledge support from the NSW Node of the Australian National Fabrication Facility. We thank Robin Cantor and STAR Cryoelectronics for sputtering the NbTiN film, and David Niepce from Low Noise Factory for discussions on circulator losses. 

\section{Author contributions}
A.V., A.K., and W.V. performed the experiments and analyzed the data. A.K., W.V. and J.J.P. designed the device. W.V. fabricated the devices. T.D. designed the device enclosures and variable thermal source. J.J.P. and A.M. supervised the project. A.V., A.K., W.V., and J.J.P. wrote the manuscript with input from all authors.   

\section{Additional information}
Online supplementary information accompanies this paper. Correspondence should be addressed to J.J.P.



\end{document}


\title{Supplementary Information: Strong Microwave Squeezing Above 1 Tesla and 1 Kelvin}

\author{Arjen Vaartjes}
\affiliation{School of Electrical Engineering and Telecommunications, UNSW Sydney, Sydney, NSW 2052, Australia}
\author{Anders Kringh\o{}j}
\affiliation{School of Electrical Engineering and Telecommunications, UNSW Sydney, Sydney, NSW 2052, Australia}
\author{Wyatt Vine}
\affiliation{School of Electrical Engineering and Telecommunications, UNSW Sydney, Sydney, NSW 2052, Australia}
\author{Tom Day}
\affiliation{School of Electrical Engineering and Telecommunications, UNSW Sydney, Sydney, NSW 2052, Australia}
\author{Andrea Morello}
\affiliation{School of Electrical Engineering and Telecommunications, UNSW Sydney, Sydney, NSW 2052, Australia}
\author{Jarryd J. Pla}
\email[]{jarryd@unsw.edu.au}
\affiliation{School of Electrical Engineering and Telecommunications, UNSW Sydney, Sydney, NSW 2052, Australia}
\date{\today}

\maketitle

\tableofcontents

\newcommand{\beginsupplement}{%
        \setcounter{table}{0}
        \renewcommand{\thetable}{S\arabic{table}}%
        \setcounter{figure}{0}
        \renewcommand{\thefigure}{S\arabic{figure}}%
     }
\beginsupplement

\section{Experimental Setup}

\begin{figure}[b!]
    \centering\includegraphics[width=0.7\textwidth]{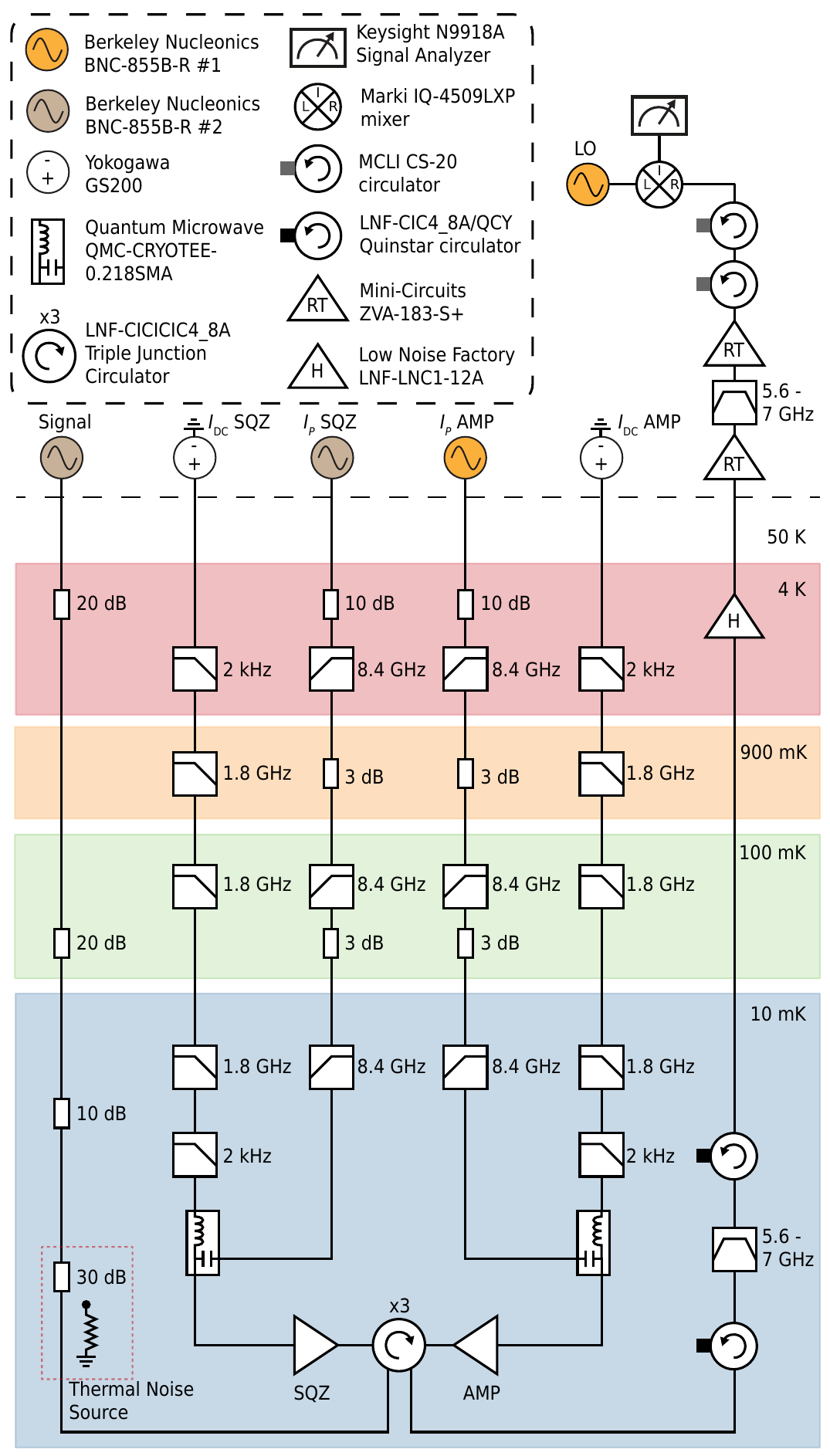}
    \caption{A wiring diagram for the commercial dilution refrigerator used in experiments.}
    \label{fig:venus_wiring}
\end{figure}

\subsection{Dilution Refrigerator Wiring}
\begin{figure}[t!]
	\centering\includegraphics[width=0.85\textwidth]{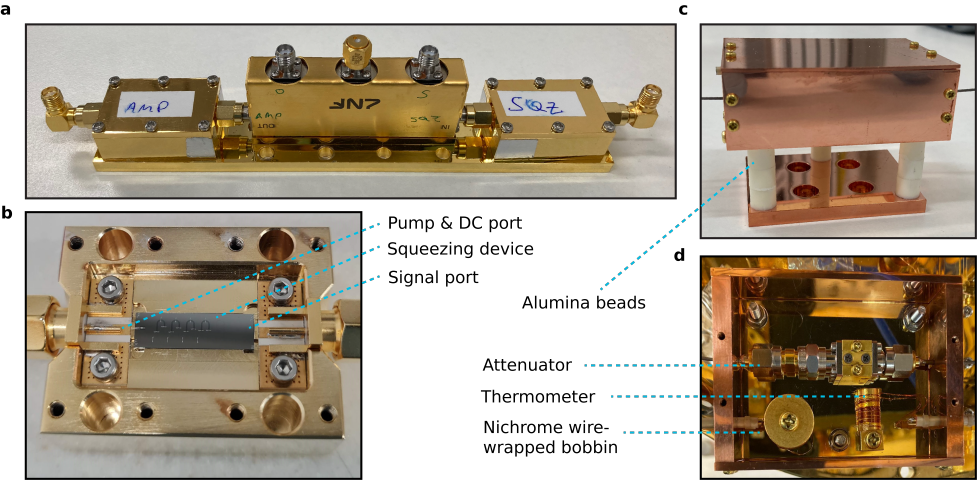}	
	\caption{Custom components used for squeezing measurements. (\textbf{a}) The two-port KIPA design allows us to connect two KIPAs with only a single circulator between them. The device enclosures and circulator are mounted to a custom baseplate to ensure alignment of the SMA connectors. (\textbf{b}) The bottom half of a KIPA enclosure with a squeezing device. (\textbf{c}), The enclosure for the thermal noise source. The box is mounted on stainless steel screws with alumina beads to prevent heat transfer from the noise source to the mixing chamber plate. (\textbf{d}) The inside of the thermal noise source, consisting of a 30~dB attenuator, a thermometer and a heater consisting of a gold-plated copper bobbin wrapped with nichrome wire.}
	\label{fig:sqz_components}
\end{figure}

A schematic depicting the lines of a commercial dilution refrigerator used in the experiments is shown in Fig.~\ref{fig:venus_wiring}. SQZ and AMP are connected to a triple-junction circulator (Low Noise Factory LNF-CICICIC4-8A).
The device enclosures and the circulator are mounted together on a custom-made gold-plated oxygen-free copper block which is thermally anchored to the mixing chamber (MXC) plate at 10~mK. This design allows the male SMA connectors of the device enclosures to connect directly to the female SMA connectors of the circulator to minimize insertion loss (Fig.~\ref{fig:sqz_components}a), and is the configuration we use for the main squeezing experiments. The circulator port prior to SQZ is used for injecting coherent signals produced via a room temperature ultra-low phase-noise microwave source (Berkeley Nucleonics BNC-845b) or thermal noise (see below). The final port of the circulator routes the signals to a HEMT at 4~K (Low Noise Factory LNF-LNC1-12A). Depending on the measurement, up to two additional room temperature amplifiers are used (Mini-Circuits ZVA-183-GX+).

A total of four lines are used to supply the pump and $\Idc$ to each KIPA, with both pairs of lines nominally identical to one another. The lines for each control signal are extensively filtered across the various temperature stages to prevent thermal noise at $\omega_0$ from reaching the KIPAs. The pump and $\Idc$ signals are combined at the MXC with a bias tee (Quantum Microwave QMC-CRYOTEE-0.218SMA) and fed into one port of the device (see below). We use home-built low-pass filters for filtering the DC lines at 4~K and at the MXC. The low-pass filter at 4~K is made from a 40~$\Omega$ length of nichrome wire wrapped around a gold-plated copper bobbin with a total of 1.88~$\mu$F of shunt capacitance, thus realizing a series RC-filter with a 2~kHz cut-off. The low-pass filter used at the MXC uses copper wire to avoid ohmic heating. These filters are enclosed in an aluminum box, with a copper base plate to enhance their thermal contact with the fridge. The inside of each box is lined with eccosorb (Laird Technologies EMI 21109145) to further attenuate signals at microwave frequencies. These homemade filters each attenuate $>50$~dB over the frequency range $\sim 10$~kHz to 30~GHz.

\subsection{Circulator Insertion Loss}

\begin{figure}
    \centering
    \includegraphics{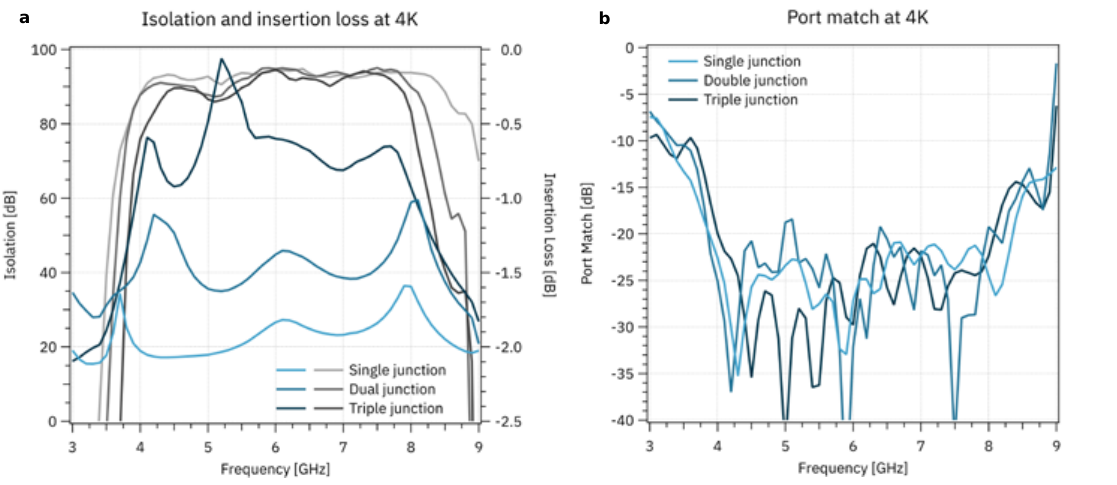}
    \caption{(\textbf{a}) Measurement of typical circulator isolation (blue lines) and insertion loss (grey lines). (\textbf{b}) Typical port match of the Low Noise Factory 4-8 GHz circulators with single, dual and triple junction(s), plotted as a function of the signal frequency. Plots supplied by Low Noise Factory.}
    \label{fig:loss_circulator}
\end{figure}

The data in Fig.~\ref{fig:loss_circulator}a depicts the typical insertion loss and isolation performance of single, double and triple junction ferrite circulators produced by the manufacturer Low Noise Factory. From this plot we estimate the insertion loss of the triple junction circulator used in our experiments to be -0.2~dB at cryogenic temperatures. The total insertion loss of the circulator can be decomposed into three primary components: insertion loss due to reflections from the connectors; loss from the internal solder joints; and loss from the garnet material that is used to provide the circulator with its nonreciprocal behavior. Here we break down the contribution of each of these elements.

From the port match plot shown in Fig.~\ref{fig:loss_circulator}b, we estimate a return loss of 22~dB at $\omega/2\pi$=6.23~GHz at a temperature of 4~K. This corresponds to an insertion loss of approximately $10\log(1-10^{-22/10})=-0.03$~dB per connector. The solder joint resistance and connector mated pair resistance contribute approximately 0.05~dB over the frequency range of 4-12~GHz per connector (private communication with Low Noise Factory). This leaves an estimated garnet dissipation loss of $0.04$~dB. If the two KIPAs and triple-junction circulator were integrated in the same enclosure, the connector and solder joint insertion losses would be eliminated. Assuming garnet dissipation as the sole source of insertion loss and the amplifier-added noise was completely mitigated, squeezing would be limited to $-20.4$~dB.

\subsection{Thermal Noise Source}
To calibrate the noise added by each KIPA we utilize a home-built variable temperature thermal noise source (TNS) (Fig.~\ref{fig:sqz_components}e). The design of the TNS is inspired by Simbierowicz, et al.~\cite{Simbierowicz2021} and consists of a 30~dB cryo-attenuator (Quantum Microwaves QMC-CRYOATT-30) that is heated by a nichrome wire wrapped around a copper bobbin. A low-temperature ruthenium oxide thermometer (Ice Oxford PKE0272) calibrated down to 30~mK is used to monitor the temperature of the TNS. The TNS is placed inside an oxygen-free copper box which is thermally isolated from the MXC via alumina beads wrapped around stainless steel screws. The thermal conductance of alumina rapidly diminishes below 3~K and thus serves to efficiently cool the noise source to this temperature. Below 3~K cooling predominantly occurs through the stainless steel mounting screws, which ensures that the noise source has sufficient thermal isolation from the MXC~\cite{Simbierowicz2021}. Superconducting NbTi cables are used to connect the heated attenuator to the circulator and input line to minimize loss and provide thermal isolation.

\section{Theory}

In this section we present the input-output theory used to model the squeezing measurements in the squeezed-state receiver setup.

\subsection{Input-Output Theory}

\subsubsection{Single KIPA in Phase-Sensitive Mode}\label{sec:iot}

For a single KIPA in phase-sensitive mode, the variance of the mode quadrature operators at its output (proportional to the noise power) can be directly related to the variance of the operators at its input via the input-output relations for the KIPA, which were derived in Reference~\cite{Parker2021}. The quadrature operators for the input propagating mode are given by

\begin{equation}
I_{\mathrm{in}} = \frac{a_{\mathrm{in}} + a^\dag_{\mathrm{in}}}{2},
\;\;\;
Q_{\mathrm{in}} = \frac{a_{\mathrm{in}} - a^\dag_{\mathrm{in}}}{2i}
,\end{equation}

\noindent where $a_{\mathrm{in}}$ and $a_{\mathrm{in}}^\dag$ are the bosonic annihilation and creation operators for the input mode, respectively. The input-output relations are then

\begin{gather}
\begin{pmatrix} I_{\mathrm{out}} \\ Q_{\mathrm{out}} \end{pmatrix} = 
\underbrace{\begin{pmatrix} A & B \\ C & D \end{pmatrix}}_{G}
\begin{pmatrix} I_{\mathrm{in}} \\ Q_{\mathrm{in}} \end{pmatrix} + 
\sqrt{\frac{\gamma}{\kappa}}
\underbrace{\begin{pmatrix} A + 1 & B \\ C & D + 1 \end{pmatrix}}_{G + 1}
\begin{pmatrix} I_{b} \\ Q_{b} \end{pmatrix}   
\label{eq:in_out_gain}
,\end{gather}

\noindent where $G$ is equal to

\begin{equation}
\begin{split}
G(\phi) = \frac{\kappa}{\Delta^2+\bar{\gamma}^2-|\zeta|^2}
\begin{pmatrix} \bar{\gamma}-|\zeta|\sin(\phi) & -|\zeta|\cos(\phi) + \Delta \\
-|\zeta_a|\cos(\phi) - \Delta & \bar{\gamma}+|\zeta|\sin(\phi)
\end{pmatrix} - 1
\end{split} 
\label{eqn_G_kipa}
,\end{equation}

\noindent and $I_{b}$ and $Q_{b}$ are the quadrature operators for the bath mode that is coupled to the KIPA. They are similarly defined as

\begin{equation}
I_{b} = \frac{b_{\mathrm{in}} + b^\dag_{\mathrm{in}}}{2},
\;\;\;
Q_{b} = \frac{b_{\mathrm{in}} - b^\dag_{\mathrm{in}}}{2i}
,\end{equation}

\noindent where $b_{\mathrm{in}}$ and $b_{\mathrm{in}}^\dag$ are the bosonic annihilation and creation operators for the bath mode, respectively,

Solving for the variance at the device output $\delta I^2_{\mathrm{out}}$, we find

\begin{gather}
\delta I^2_{\mathrm{out}} = A^2(\delta I_{\mathrm{in}}^2) + B^2 (\delta Q_{\mathrm{in}}^2) + \frac{\gamma}{\kappa}\left[ (A+1)^2 (\delta I_{b}^2)+ B^2(\delta Q_{b}^2)\right]
\label{eq:Isq_out}
.\end{gather}

\noindent In Eq.~\ref{eq:Isq_out} we have made use of the fact that the orthogonal quadratures of the input field are uncorrelated (i.e. $\left<{I_{\mathrm{in}},Q_{\mathrm{in}}}\right>=0$) as are the input and bath fields (e.g. $\left<{I_{\mathrm{in}},Q_{b}}\right>=0$). Note that the variance of the $I$ and $Q$ quadratures are in general linked to one another, however, if $\Delta=\cos(\phi)=0$ they become linearly independent. In this case,

\begin{gather}
\delta I^2_{\mathrm{out}} = G_k (\delta I_{\mathrm{in}}^2) + (G_k-1)n_k
\label{eqn_input_output_X2_kipa}
,\end{gather}

\noindent where $G_k=A^2$ is the power gain of the KIPA. When $G_k > 1$, which occurs when $\phi=3\pi/2$, the KIPA anti-squeezes the noise along $I$. When $0 < G_k < 1$, which occurs when $\phi=\pi/2$, the KIPA squeezes the noise along $I$. $n_k$ is the total number of noise photons added by the KIPA to the $I$ quadrature and is equal to

\begin{equation}
n_k = \frac{\gamma}{\kappa}\frac{A+1}{A-1}\delta {I^2_{b}} = \frac{\gamma}{\kappa}\frac{A+1}{A-1}\left(\frac{1}{4} + n_{k,\mathrm{th}}\right)
\label{eqn_noise_amp_defn}
,\end{equation}

\noindent where $n_{k,\mathrm{th}}$ is the noise added per quadrature of $b_{\mathrm{in}}$ measured in excess of vacuum, in units of photons.

Equation~\ref{eqn_input_output_X2_kipa} is intuitive in that by deactivating the KIPA, which is equivalent to setting $G_k=1$, the input mode simply reflects off the KIPA without any noise being added. Note that taken in isolation, Eq.~~\ref{eqn_noise_amp_defn} seems to suggest that the noise added by the KIPA is negative when $A < 1$, i.e. when the KIPA is configured to squeeze. However, we see that this is accounted for in Eq.~\ref{eqn_input_output_X2_kipa} because the term $G_k - 1 < 1$, such that the noise added to the $I$-quadrature is always greater than zero whenever $G_k \neq 1$.

\begin{figure}[h]
    \centering\includegraphics[width=0.6\textwidth]{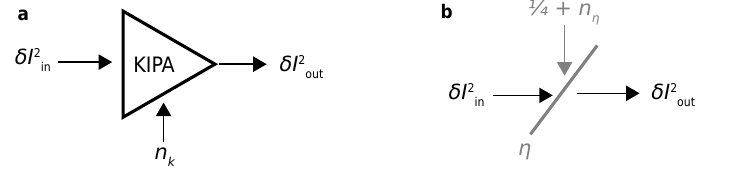}
    \caption{(\textbf{a}) Input-output model of a single KIPA, corresponding to Eq.~\ref{eq:in_out_gain}. (\textbf{b}) Input-output model of a beamsplitter, corresponding to Eq.~\ref{eq:loss}, which is used to describe the finite insertion loss of components in the setup.}
    \label{fig:enter-label}
\end{figure}

\subsubsection{Input-Output Relation for the Loss Between Components}
\FloatBarrier

We can model the loss between two components in the setup as a beamsplitter with transmission efficiency $\eta$. The beamsplitter both attenuates the input mode and mixes in noise from the bath. The noise of the bath, stated in terms of noise photons, is given by $\delta I^2_{\eta} = 1/4 + n_\eta$, where $n_\eta = [\exp(\hbar\omega_0/k_BT_{\eta})-1]^{-1}/2$ is the Bose-Einstein occupation (per quadrature) for a field with a frequency $\omega_0$ and at a temperature $T_{\eta}$. The input-output relation for the beamsplitter is then

\begin{equation}
\begin{split}
    \delta I^2_{\mathrm{out}} & = \eta (\delta I^2_{\mathrm{in}}) + (1-\eta) (\delta I^2_{\eta}) \\ 
    & = \eta (\delta I^2_{\mathrm{in}}) + (1-\eta) \left(\frac{1}{4} + n_{\eta}\right)
\end{split}
\label{eq:loss}
.\end{equation}

\subsubsection{Input-Output Relation for the Full Experimental Setup}

\begin{figure}[h]
    \centering\includegraphics{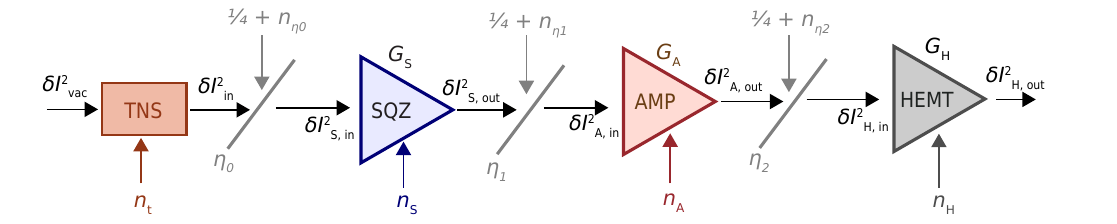}
    \caption{Input-output diagram containing a Thermal Noise Source (TNS), two KIPAs (SQZ and AMP) and a HEMT amplifier. Each component is linked with a non-ideal transmission modelled as a beamsplitter with transmission efficiency $\eta_{i}$}
    \label{fig:input-output}
\end{figure}

\FloatBarrier
\noindent The input-output relations for the full setup can be derived by linking the input-output relations for each discrete component and accounting for the finite insertion loss between them. An input-output diagram of the direct squeezing setup (Fig.~1 of the main text) is shown in Fig.~\ref{fig:input-output}. It is comprised of a thermal noise source (TNS), two KIPAs, and a High Electron Mobility Transistor (HEMT) amplifier, all connected in series via a circulator. The reader is referred to Fig.~\ref{fig:input-output} for the definition of each term in the following input-output relations. We ignore the room temperature amplifiers in our setup (Fig.~\ref{fig:venus_wiring}) since $G_Hn_H$ far exceeds the noise added by these devices, or in other words, the noise in our measurements is dominated by sources within the dilution refrigerator. 

We build a series of coupled linear equations, going from left to right in Fig.~\ref{fig:input-output}:

\begin{gather}
\delta I^2_{\mathrm{in}} = \delta I^2_{\mathrm{vac}} + n_\mathrm{t} = \frac{1}{4} + n_\mathrm{t}
\label{eq:input_output_in}
\\
\delta I^2_{\mathrm{S, in}} = \eta_0 (\delta I^2_{\mathrm{in}}) + (1-\eta_0)\left(\frac{1}{4} + n_{\eta_0} \right)
\label{eq:input_output_eta_0}
\\
\delta I^2_{S,\mathrm{out}} = \Gsqz(\delta I_{\mathrm{S, in}}^2) + (\Gsqz-1)n_{\mathrm{S}}
\label{eq:input_output_SQZ}
\\
\delta I^2_{\mathrm{A, in}} = \eta_1(\delta I^2_{\mathrm{S, out}}) + (1-\eta_1)\left(\frac{1}{4} + n_{\eta_1} \right)
\label{eq:input_output_eta_1}
\\
\delta I^2_{\mathrm{A, out}} = \Gamp(\delta I_{A,\mathrm{in}}^2) + (\Gamp-1)n_\mathrm{A}
\label{eq:input_output_AMP}
\\
\delta I^2_{\mathrm{H, in}} = \eta_2(\delta I^2_{\mathrm{A, out}}) + (1-\eta_2)\left(\frac{1}{4} + n_{\eta_2} \right)
\label{eq:input_output_eta_2}
\\
\delta I^2_{\mathrm{H, out}} = G_\mathrm{H}(\delta I_{\mathrm{H, in}}^2) + (G_\mathrm{H}-1)n_\mathrm{H}
\label{eq:input_output_HEMT}
.\end{gather}

\noindent In Eq.~\ref{eq:input_output_in}, $n_t=[\exp(\hbar\omega/k_B\Th)]^{-1}/2$ is the thermal noise added by the TNS to the $I$ quadrature. In Eq.~\ref{eq:input_output_HEMT}, $n_\mathrm{H}  = 1/4 + n_{H,\mathrm{th}}$ is the total noise added by the HEMT to the $I$-quadrature, where $n_{H,\mathrm{th}}$ is the noise added by the HEMT in excess of vacuum.

Combining equations~\ref{eq:input_output_in}-\ref{eq:input_output_HEMT} we create a single input-output relation for the experimental setup

\begin{gather}
\begin{split}
\delta I^2_{\mathrm{H, out}} & = G_\mathrm{H}\Gamp\Gsqz\eta_0\eta_1\eta_2\left(\frac{1}{4} + n_\mathrm{t} \right) \\
& + G_\mathrm{H}\Gamp\Gsqz(1-\eta_0)\eta_1\eta_2 \left(\frac{1}{4} + n_{\eta 0}\right) \\
& + G_\mathrm{H}\Gamp(\Gsqz-1)\eta_1\eta_2n_\mathrm{S} \\
& + G_\mathrm{H}\Gamp(1-\eta_1)\eta_2\left(\frac{1}{4} +  n_{\eta 1}\right) \\
& + G_\mathrm{H}(\Gamp - 1)\eta_2 \na \\
& + \nsys,
\end{split}
\label{eq:input_output_all}
,\end{gather}

\noindent where for conciseness we group together the system noise $\nsys$ (referred to the output of the HEMT amplifier), which is equal to

\begin{gather}
\begin{split}
\nsys & = G_\mathrm{H} (1-\eta_2)\left(\frac{1}{4}+n_{\eta 2}\right) + (G_\mathrm{H} -1)n_\mathrm{H}
\end{split}
.\end{gather}

\noindent The input-output relation in Eq.~\ref{eq:input_output_all} is the theoretical framework for all of the experiments in the main text. In the next sections we apply it to determine what limits our measurements of direct squeezing, to define a procedure for calibrating the noise added by SQZ and AMP, and for explaining the thermal state squeezing experiments.

\subsection{Limitations of the Direct Squeezing Measurements}

As described in the main text, we make direct measurements of vacuum squeezing by comparing the power of the noise along $I$ when both SQZ and AMP are activated and when only AMP is activated. Using Eq.~\ref{eq:input_output_all}, this measurement protocol is equal to

\begin{equation}
    S
    = \frac{
    \left.\delta I_{\mathrm{H,out}}^2\right|_{G_s\ne1, G_a\ne1, n_t=0}
    }
    {
    \left.\delta I_{\mathrm{H,out}}^2\right|_{G_s=1, G_a\ne1, n_t=0}
    } :=
    \frac{\ISA}{\IA}
    \label{eq:squeezing_measurement}
.\end{equation}

We assume that the bath temperatures for the losses $\eta_0$ and $\eta_1$ are both of the order of $\Tmc \approx 10$~mK, which ensures that $n_{\eta_0} = n_{\eta_1} \approx 0$. The numerator of Eq.~\ref{eq:squeezing_measurement} then becomes

\begin{equation}
    \ISA
    =
    G_\mathrm{H} \Gamp \eta_2 
    \left[ 
    (\Gsqz - 1)\eta_1 \left(\frac{1}{4} + n_\mathrm{S}
    \right)
    + \frac{1}{4} + n_\mathrm{A}\left(1-\frac{1}{\Gamp}\right)
    \right] + \nsys,
\end{equation}

\noindent and the denominator becomes

\begin{equation}
    \IA
    =
    G_\mathrm{H} \Gamp \eta_2 
    \left[ 
    \frac{1}{4} + n_\mathrm{A}\left(1-\frac{1}{\Gamp}\right)
    \right] + \nsys
.\end{equation}

\noindent Combining these expressions we find

\begin{equation}
    S = 1 - \frac{\eta_1 (1-\Gsqz)(1/4-|n_\mathrm{S}|)}{1/4+(1-1/\Gamp)n_\mathrm{A} + \nsys/(G_\mathrm{H} \Gamp \eta_2)}
    \label{eq:squeezing_final}
,\end{equation}

\noindent where we have taken the absolute value of $n_\mathrm{S}$ in the second line to clarify that when $\Gsqz<1$, $n_\mathrm{S} < 0$ (see Sec.~\ref{sec:iot}). Provided $|n_\mathrm{S}| < 1/4$, we see that Eq.~\ref{eq:squeezing_final} predicts that activating SQZ should reduce the noise measured at the output of the system.

In the limit of high AMP gain ($\nsys/\Gamp\Gh\eta_2\ll1/4$ and $\Gamp \gg 1$), i.e. when the system noise referred to the input of AMP is sufficiently smaller than the vacuum fluctuations, this expression reduces to Eq.~1 in the main text. 
Equation~\ref{eq:squeezing_final} exposes the two main experimental limitations in the direct measurement of squeezing: the insertion loss between SQZ and AMP, $\eta_1$, and the noise added by the amplifiers, $n_\mathrm{S}$ and $n_\mathrm{A}$. 

\subsection{Amplifier Noise Calibration}
\label{SI:noise_temp_theory}

We calibrate the noise added by both SQZ and AMP by using the TNS to vary the thermal occupation of the input state $n_t$. Crucially, the KIPAs remain stable at $\sim 10$~mK. The noise added by the KIPAs $n_\mathrm{S}$ and $n_\mathrm{A}$ can then be found through the relation between the power of the noise measured at the output of the system and $n_t$. 

To find $n_\mathrm{S}$ and $n_\mathrm{A}$ we measure $\delta I^2_{\mathrm{H, out}}$ as a function of $n_\mathrm{t}$ in three different scenarios: SQZ activated ($\Gamp = 1$), AMP activated ($\Gsqz = 1$), and both SQZ and AMP deactivated ($\Gamp, \Gsqz = 1$). For each scenario, using Eq.~\ref{eq:input_output_all} we obtain a linear relation between $\delta I^2_{\mathrm{H, out}}$ and $n_\mathrm{t}$ and find slopes that are given by

\begin{gather}
s_{\mathrm{S}} := \left.\frac{d\delta I^2_{\mathrm{H, out}}}{d n_\mathrm{t}}\right|_{\Gamp=1} = \Gh\Gsqz \eta_0\eta_1\eta_2
, \label{slope_sqz}
\\
s_\mathrm{A} := \left.\frac{d\delta I^2_{\mathrm{H, out}}}{d n_\mathrm{t}}\right|_{\Gsqz=1} = \Gh\Gamp \eta_0\eta_1\eta_2
, \label{slope_amp}
\\
s_\mathrm{off} := \left.\frac{d\delta I^2_{\mathrm{H, out}}}{d n_\mathrm{t}}\right|_{\Gsqz=\Gamp=1} = \Gh \eta_0\eta_1\eta_2 \label{slope_hemt}
,\end{gather}

\noindent and with y-intercepts

\begin{gather}
\begin{split}
    y_\mathrm{S} := \left.\delta I^2_{\mathrm{H, out}}\right|_{\Gamp=1,n_\mathrm{t}=0} = & \frac{\Gh\Gsqz\eta_0\eta_1\eta_2}{4} + \Gh\Gsqz(1-\eta_0)\eta_1\eta_2\left(\frac{1}{4} + n_{\eta 0}\right) \\
& + \Gh(\Gsqz-1)\eta_1\eta_2 \ns + \Gh(1-\eta_1)\eta_2\left(\frac{1}{4} + n_{\eta 1}\right) + \nsys
\label{y_intercept_sqz}
\end{split}
,\\
\begin{split}
y_\mathrm{A} := \left.\delta I^2_{\mathrm{H, out}}\right|_{\Gsqz=1,n_\mathrm{t}=0} = & \frac{\Gh\Gamp\eta_0\eta_1\eta_2}{4} + \Gh\Gamp(1-\eta_0)\eta_1\eta_2 \left(\frac{1}{4} + n_{\eta 0}\right) \\
& + \Gh\Gamp(1-\eta_1)\eta_2\left(\frac{1}{4} +  n_{\eta 1}\right) + \Gh(\Gamp - 1)\eta_2 \na + \nsys
\label{y_intercept_amp}
\end{split}
,\\
\begin{split}
y_\mathrm{off} := \left.\delta I^2_{\mathrm{H, out}}\right|_{\Gsqz=\Gamp=1,n_\mathrm{t}=0} = & \frac{\Gh\eta_0\eta_1\eta_2}{4} + \Gh(1-\eta_0)\eta_1\eta_2 \left(\frac{1}{4} + n_{\eta 0}\right) \\
& + \Gh(1-\eta_1)\eta_2\left(\frac{1}{4} + n_{\eta 1}\right) + \nsys
\label{y_intercept_hemt}
\end{split}
.\end{gather}

\noindent We now define a new parameter $\delta \Tilde{I}^2_\mathrm{k} := \delta I^2_{\mathrm{k}} - \delta I^2_\mathrm{off}$, where $k \in \{S, A\}$, i.e. the difference between the noise measured with SQZ or AMP on with the noise measured when both SQZ and AMP are deactivated.
For $\delta \Tilde{I}^2_\mathrm{S}$, we obtain a y-intercept of

\begin{gather}
    \begin{split}
        \Tilde{y}_\mathrm{S} = 
        y_\mathrm{S} - y_\mathrm{off} = & \frac{\Gh(\Gsqz-1)\eta_0\eta_1\eta_2}{4} + \Gh(\Gsqz-1)(1-\eta_0)\eta_1\eta_2\left(\frac{1}{4} + n_{\eta 0}\right) \\
& + \Gh(\Gsqz-1)\eta_1\eta_2 \ns \\
& = \Gh(\Gsqz-1)\eta_1\eta_2\left[\frac{1}{4} + (1-\eta_0)n_{\eta 0} + \ns \right] 
    \end{split}
\end{gather}

\noindent and a slope of

\begin{equation}
\Tilde{s}_\mathrm{S} = 
    s_{\mathrm{S}} - s_\mathrm{off} = \Gh(\Gsqz-1)\eta_0\eta_1\eta_2.
\end{equation}

\noindent Similarly, for the case where AMP is activated, we find a y-intercept of

\begin{gather}
    \begin{split}
        \Tilde{y}_\mathrm{A} = 
        y_\mathrm{A} - y_\mathrm{off} = & \frac{\Gh(\Gamp-1)\eta_0\eta_1\eta_2}{4} + \Gh(\Gamp-1)(1-\eta_0)\eta_1\eta_2 \left(\frac{1}{4} + n_{\eta 0}\right) \\
& + \Gh(\Gamp-1)(1-\eta_1)\eta_2\left(\frac{1}{4} + n_{\eta 1}\right) + \Gh(\Gamp - 1)\eta_2 \na \\
& = \Gh(\Gamp-1)\eta_2\left[\frac{1}{4} + (1-\eta_0)n_{\eta 0} + (1-\eta_1) n_{\eta 1} + \na \right]
    \end{split}
\end{gather}

\noindent and a slope of

\begin{equation}
    \Tilde{s}_\mathrm{A} = 
    s_{\mathrm{A}} - s_\mathrm{off} = \Gh(\Gamp-1)\eta_0\eta_1\eta_2.
\end{equation}

Dividing the y-intercept and the slope allows us to extract the noise added by SQZ and AMP, $\ns$ and $\na$, respectively. For `SQZ on', we obtain

\begin{equation}
\frac{\Tilde{y}_\mathrm{S}}{\Tilde{s}_{\mathrm{S}}} = \frac{1}{\eta_0}\left[\frac{1}{4} + (1-\eta_0)n_{\eta 0} + \ns\right],
    \label{eq:noise_temp_sqz}
\end{equation}

\noindent and for `AMP on' we obtain

\begin{equation}
    \frac{\Tilde{y}_\mathrm{A}}{\Tilde{s}_{\mathrm{A}}} = \frac{1}{\eta_0\eta_1}\left[\frac{1}{4} + (1-\eta_0)n_{\eta 0} + (1-\eta_1) n_{\eta 1} + \na \right].
    \label{eq:noise_temp_amp}
\end{equation}

Using this model we determine the noise added by SQZ and AMP in Section~\ref{SI:noise_temp_meas}.

\subsection{Squeezing of Thermal States}

For the experiments at elevated temperatures, described in Figs.~4\textbf{d-g} of the main text, the setup is modified such that SQZ and the circulator are thermally connected to the TNS. The temperatures for the beamsplitters $\eta_0$ and $\eta_1$ are assumed to be the same as the temperature of the input signal, i.e., $n_{\eta_0}=n_{\eta_1}=n_\mathrm{t}$. In this case, Eq.~\ref{eq:input_output_all} simplifies to

\begin{gather}
\begin{split}
\delta I^2_{\mathrm{H, out}} & = \Gh\Gamp\eta_2\left[(\Gsqz - 1)\eta_1(\frac{1}{4} + n_\mathrm{t} + \ns) + 1/4 + n_t + \left(1-\frac{1}{\Gamp}\right)\na\right] + \nsys.
\end{split}
\label{eq:input_output_all_v2}
\end{gather}

\noindent Equation~\ref{eq:squeezing_final} can similarly be re-written as

\begin{equation}
    S = \frac{\ISA(n_\mathrm{t})}{\IA(n_\mathrm{t})} = 1 - (1 - \Gsqz)\eta_1\frac{1/4 + n_\mathrm{t} + \ns}{1/4 + n_\mathrm{t} + (1-1/\Gamp)\na + \nsys/\Gh\Gamp\eta_2}
.\end{equation}

\noindent In the high AMP gain limit ($\nsys/\Gamp\Gh\eta_2\ll1/4$ and $\Gamp \gg 1$), this expression reduces to

\begin{equation}
    S = \frac{\ISA(n_\mathrm{t})}{\IA(n_\mathrm{t})} = 1 - (1 - \Gsqz)\eta_1\frac{1/4 + n_\mathrm{t} + \ns}{1/4 + n_\mathrm{t} + \na}.
    \label{eq:squeezing_thermal}
\end{equation}

We fit the data in Supplementary Fig.~\ref{fig:thermal_squeezing_fit} to this expression to extract $\eta_1$ and the noise added by SQZ when operated in squeezing mode $\ns^{\mathrm{sq}}$.

\section{Device Fabrication, Design, and Packaging}
\FloatBarrier

\subsection{Fabrication}
The two KIPAs in our experiment are nominally identical in their design and fabrication. They are fabricated from a 15~nm thin film of NbTiN deposited on high-resistivity silicon ($>10$~k$\Omega\cdot$cm). They are patterned with a single step of electron beam lithography and subsequently etched with a CF\textsubscript{4}:Ar plasma. The kinetic inductance of the film was calibrated to be $L_{k,0}=17.8~$pH/$\square$ by fabricating a capacitively coupled quarter-wavelength ($\lambda/4$) coplanar waveguide (CPW) resonator, and matching its measured resonant frequency with electromagnetic simulations of the device using the software package Sonnet.  

\subsection{Design}
\begin{figure*}[t!]
\includegraphics[width=0.8\textwidth]{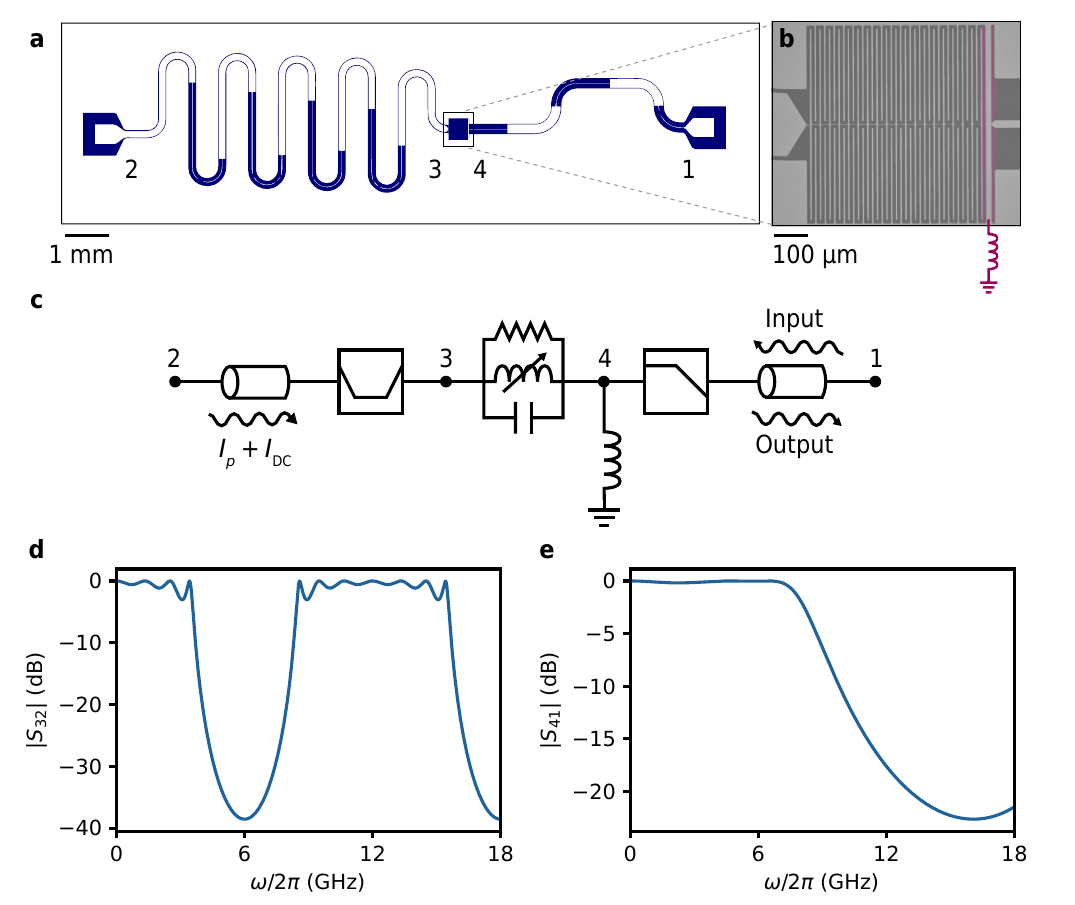}
\caption{\textbf{Design of the KIPAs.} (\textbf{a}) The device layout. The right port (1) connects to a low-pass stepped-impedance filter (LP-SIF). The left port (2) connects to a band-stop stepped impedance filter (BS-SIF). Both the LP-SIF and BP-SIF are constructed from alternating high $Z_h$- and low $Z_l$-impedance CPW segments. The half-wavelength resonator is constructed from a CPW with a large interdigitated capacitance (IDC) to ground and is positioned in the centre of the device. (\textbf{b}) An optical image of the resonator. The shunt inductance controlling the coupling quality factor $Q_c$ is highlighted in purple. (\textbf{c}) Simplified circuit diagram of the device. (\textbf{d}) The scattering parameter $|S_{32}|$ of the BS-SIF, numerically calculated using ABCD matrices. (\textbf{e}) The scattering parameter $|S_{41}|$ of the LP-SIF, numerically calculated using ABCD matrices.}
\label{fig:design}
\end{figure*}

The devices are designed to have a half-wavelength ($\lambda/2$) resonator connected to two ports. One of the ports connects directly to the resonator via a band-stop stepped impedance filter (BS-SIF, left side in Fig.~\ref{fig:design}a), while the other is inductively coupled to the resonator (Fig.~\ref{fig:design}b) via a low-pass stepped impedance filter (LP-SIF, right side in Fig.~\ref{fig:design}a). A simplified circuit model of the device is presented in Fig.~\ref{fig:design}c. We employ a two-port design because it allows the pump tone, DC current $\Idc$ and resonant tones (including squeezed states) to be independently routed off-chip. This is essential for squeezing experiments because it allows for the connection of SQZ and AMP in series with only a single intervening circulator. This ensures a minimum amount of insertion loss, which would otherwise degrade the squeezed states as they travel between SQZ and AMP.

The resonator contains a dense interdigitated capacitance (IDC) to ground (Fig.~\ref{fig:design}b), which lowers its impedance and reduces its physical size, relative to a CPW design. Lowering the impedance of the resonator improves the performance of the amplifier by increasing the pump current for a given power. This results in a greater modulation of the kinetic inductance and therefore higher gain.


The BS-SIF is constructed from a total of nine $\lambda/4$ CPW segments, with alternating low-impedance ($Z_l$, five segments) and high-impedance ($Z_h$, four segments)~\cite{Bronn2015}. The resonator is directly connected to a $Z_l$ segment, which in combination with the inductive shunt on its opposite end, results in electrical boundary conditions that give rise to a $\lambda/2$ resonant mode. The purpose of the BS-SIF is to pass both a DC current $\Idc$ and pump current $I_\mathrm{p}$ to the resonator, while strongly attenuating at the resonance frequency $\omega_0/2\pi$.

The LP-SIF is constructed from a total of five segments of CPW with alternating $Z_h$ (three segments) and $Z_l$ (two segments). The LP-SIF is introduced to filter out the pump tone at $\omega_p\approx2\omega_0$ in order to limit crosstalk between SQZ and AMP (see Fig.~\ref{fig:design}d). It is implemented on chip using the NbTiN film to minimize insertion loss. Further, the LP-SIF is designed to be 50~$\Omega$-matched at $\omega_0/2\pi$ to minimize reflections from the signal port.

The BS-SIF and LP-SIF are designed using ABCD matrices~\cite{Pozar2012} and electromagnetic simulations (Sonnet). For each segment, we simulate CPW waveguides with centre conductor width $W$ and a gap to ground width $G$, from which we extract the effective dielectric constant $\epsilon_r$ and characteristic impedance $Z$ (see Table~\ref{table:design parameters}). This allows us to relate the electrical and physical lengths of each segment, so that a numerical model of each filter can be constructed using ABCD matrices. The numerical calculation of the BS-SIF transmission response is shown in Fig.~\ref{fig:design}d. It demonstrates a strong attenuation for frequencies near $\omega_0$, and pass bands at both DC and the pump frequency. The numerical calculation of the LP-SIF filter response is shown in Fig.~\ref{fig:design}e. It contains the desired features of minimal attenuation around $\omega_0/2\pi\approx6$~GHz followed by a sharp roll-off, which yields $\sim$20~dB attenuation at $\omega_p/2\pi\approx12$~GHz. 

The resonator is coupled to the LP-SIF via a shunt inductance, corresponding to two electrically-parallel fingers that connect to the ground plane of the device (Fig.~\ref{fig:design}b). The inductive coupling provides a path to ground for the DC current $\Idc$, while simultaneously allowing control of the coupling quality factor over several orders of magnitude \cite{Bothner2013}. The dimensions of the fingers ($12~\mu$m wide and $248~\mu$m long) were designed based on electromagnetic simulations of the device with a targeted coupling quality factor $Q_c = 200$, in close agreement with the measured results, see Fig.~\ref{fig:q example}. This ensures that the KIPA is strongly over-coupled ($Q_i \gg Q_c$) when measured in reflection via this port.

\begin{table}
\begin{center}
\begin{tabular}{ |p{2.2cm}||p{2.2cm}|p{2.2cm}|p{2.2cm}| p{2.2cm}| p{2.2cm}|}
\hline
 Parameter & $Z_h$ (BS-SIF) & $Z_l$ (BS-SIF) & $Z_h$ (LP-SIF) & $Z_l$ (LP-SIF) & IDC Resonator\\
 \hline
 $Z$~$(\Omega)$  & 131.1 & 38.7 &  	137.0 & 34.3 & 101.7\\
 $\epsilon_r$ &  14.7	& 12.0 & 14.1 & 11.8 & 3237.6\\
 $W$~$(\mu$m) & 20 & 160 &  20 & 214 & 1\\
 $G$~$(\mu$m)  & 80 & 10 &  105 & 8 & 10\\
 length~$(\mu$m) &  3264 & 3607 & 680, 1214, 680 & 1692, 1692 & 425\\
 \hline
\end{tabular}
\caption{\label{table:design parameters} Design and simulated parameters of the various KIPA components. The resonator has an interdigitated capacitance (IDC) to ground consisting of 19 fingers with length $241~\mu$m, width $4.3~\mu$m and gap $7.7~\mu$m. }
\end{center}
\end{table}

\subsection{Packaging}
\FloatBarrier

The KIPAs are mounted in separate device enclosures which are machined from gold-plated oxygen-free copper (Figs.~\ref{fig:sqz_components}b,c). The devices sit in a trench on top of a $500~\mu$m thick piece of sapphire. The sapphire acts as a low-loss dielectric spacer to ensure that the resonant mode is sufficiently separated from metallic enclosure, whilst still maintaining good thermalization. The lid of the enclosure is designed so that the cavity sits within a 3D-waveguide whose cut-off frequency is 8~GHz, which is above the design of resonator ($\omega_0/2\pi=6$~GHz). To ensure there are no gaps which might compromise the integrity of the waveguide mode, 1.5~cm long grooves are made in the enclosure $300~\mu$m away from the long edge on either side of the device. In the grooves we place a 200~$\mu$m diameter indium wire which seals the waveguide when compressed by the enclosure lid. The device is grounded to the enclosure with wire bonds from the ground plane to the enclosure. Additional wire bonds are added to equalize the ground planes at each impedance step in the BS-SIF and LP-SIF to prevent the excitation of slotline modes.

The devices are bonded to PCBs made from gold-plated copper on Rogers RO3035 laminate. The dielectric constant of this material is known to have excellent thermal stability, which is important to ensure it remains impedance matched when cooled to low temperatures. The PCBs were designed to have the minimum length necessary to facilitate the PCB-mounted SMA connector and wire bonding.

\section{Characterization of the KIPAs}

In this section, we compare the operation and performance of the two KIPAs used in our experiments.

\subsection{Reflection Measurement}

\begin{figure*}[t!]
\includegraphics[width=0.7\textwidth]{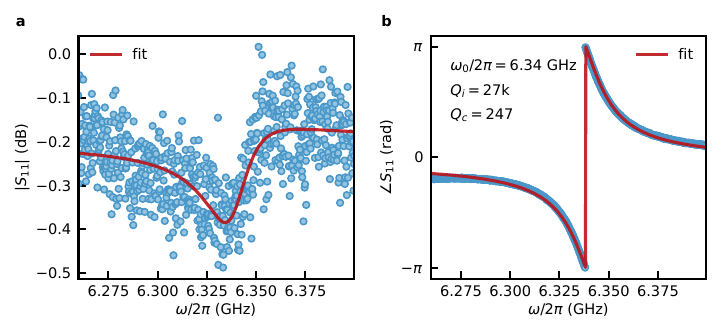}
\caption{\textbf{Reflection measurement of SQZ.} (\textbf{a,b}) Magnitude $|S_{11}|$ (\textbf{a}) and phase $\angle S_{11}$ (\textbf{b}) response of SQZ measured as a function of probe frequency $\omega$ in reflection (S11) with a vector network analyzer. The solid red lines correspond to a fit of the data in the complex plane, from which we obtain $\omega_0$, $Q_i$, and $Q_c$. The measurement is performed at $\Idc=0$ and $B_\parallel=0$.}
\label{fig:q example}
\end{figure*}

To extract the resonance frequency $\omega_0/2\pi$, internal quality factor $Q_i$ and the coupling quality factor $Q_c$ of the each KIPA, we perform a reflection measurement (measured from port 1 in Figs.~\ref{fig:design}a,c) of the magnitude $|S_{11}|$ and phase response $\angle S_{11}$ of the resonator. We then extract $\omega_0$, $Q_i$ and $Q_c$ from a combined fit~\cite{Probst2015}. An example of this using SQZ is shown in Figs.~\ref{fig:q example}a,b.

\subsection{Frequency Tunability}
Both KIPAs exhibit closely-matched resonance frequencies at zero current, with SQZ having $\omega_0/2\pi(\Idc=0)=6.341$~GHz and AMP having $\omega_0/2\pi(\Idc=0)=6.378$~GHz. In Fig.~2a of the main text we demonstrate that $\omega_0/2\pi$ of SQZ and AMP can be tuned by up to 215~MHz and 290~MHz respectively, using a DC current $\Idc$. The $\Idc$ enhances the kinetic inductance of the device according to the relation~\cite{Zmuidzinas2012}

\begin{equation}
L_k(\Idc) = L_{k,0}\left(1 + \frac{\Idc^2}{I_*^2} + \mathcal{O}(I_*^4)\right)
,\end{equation}

\noindent where $L_{k,0}$ is the kinetic inductance at zero current and $I_*$ is a constant that dictates the strength of the non-linearity. The frequency tunability of these devices is ultimately limited by their critical current $I_c$, which was found to be $1.22~$mA for SQZ and $1.35~$mA for AMP. $I_*$ can be determined by fitting $\omega_0(\Idc)$ with the expression~\cite{Parker2021}

\begin{equation}
    \omega_0(\Idc) = \omega_0(0)\left(1 - \frac{\Idc^2}{2I_*^2}\right)
    \label{eq:resonance_frequency}
,\end{equation}

\noindent and was found to be $I_*=5.1~$mA for both SQZ and AMP.

\subsection{Gain, Bandwidth and Independent Vacuum Squeezing}

\begin{figure*}[t!]
\includegraphics[width=\textwidth]{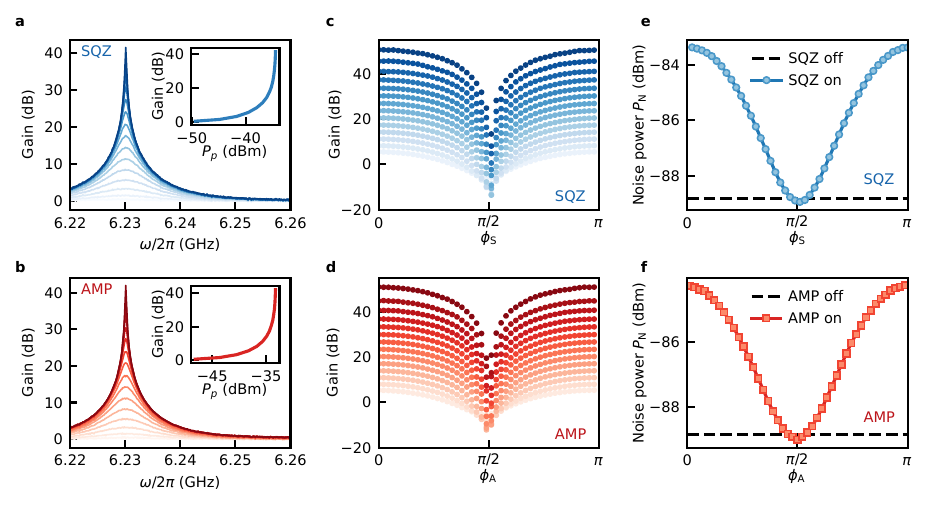}
\caption{\textbf{Gain and independent vacuum squeezing of the KIPAs.} (\textbf{a,b}) Non-degenerate gain of SQZ (\textbf{a}) and AMP (\textbf{b}) measured as a function of probe frequency $\omega/2\pi$. Increasing the opacity corresponds to increasing the pump power from $P_p=-49$ to $-34$~dBm for SQZ and from $P_p=-48$ to $-33$~dBm for AMP. Inset: Extracted maximum gain as a function of $P_p$. (\textbf{c,d}) Degenerate gain measured as a function of pump phase. The pump powers match those used in panels a and b. (\textbf{e,f}) The noise power measured along the $I$-quadrature $\PN$ as a function of pump phase. The measurements were taken with only one of SQZ or AMP on at a time. In both panels, the black dashed line corresponds to a measurement taken with both KIPAs off, and corresponds to a reference power level for when vacuum noise is being sent from the MXC plate. Vacuum squeezing occurs for pump phases about $\pi/2$, where the measured power drops below the reference level. For all panels, the devices were tuned to have frequency $\omega_0/2\pi=6.23~$GHz and the pump frequency was set to $\omega_p/2\pi=2\omega_0/2\pi=12.46$~GHz. Panels a,c,e are re-plots of the data presented in Fig.~2 of the main text.}
\label{fig:compare amp gain}
\end{figure*}

We compare the non-degenerate gain of the two KIPAs by tuning them to the same resonance frequency $\omega_0/2\pi=6.23~$GHz with a DC current, pumping them with a microwave tone with frequency $\omega_p=2\omega_0$, and measuring the scattering parameter $S_{11}$ using a vector network analyzer, one device at a time. Figures~\ref{fig:compare amp gain}a,b show the corresponding gain curves for SQZ and AMP measured as a function of pump power $P_p$. A baseline was subtracted from the $S_{11}$ measurements, which was obtained by shifting the KIPAs' resonant frequencies outside of the measurement range with a DC current. The two devices perform in a near-identical manner, with SQZ and AMP achieving a peak non-degenerate gain of 41.5~dB and 42.0~dB, respectively. The gain bandwidth products for SQZ and AMP are found to be $17$~MHz and $15$~MHz, respectively.

To compare the degenerate gain of the KIPAs, we amplify coherent states with frequency $\omega_0$ by pumping the devices with tones of frequency $\omega_p=2\omega_0$ and variable phase (Figs.~\ref{fig:compare amp gain}c,d). SQZ and AMP achieve nearly identical maximal degenerate amplification (deamplification) of 50.5~dB and 50.9~dB (-13.6~dB and -12.2~dB), respectively. We attribute the large asymmetry in amplification and deamplification to be predominantly the result of the course step size in the phase used in this measurement.

In the absence of a probe signal and with the thermal noise source turned off, the field input to SQZ (or AMP when SQZ is off) corresponds to a vacuum state. We make direct measurements of vacuum squeezing with each amplifier independently by measuring the phase-dependent noise power $\PN$ along the $I$-quadrature when one KIPA is on and the other is off, and comparing it to a measurement where both KIPAs are off (Figs.~\ref{fig:compare amp gain}e,f). SQZ and AMP are independently capable of reducing the noise below the reference vacuum level by -0.14~dB and -0.16~dB, respectively. Because these measurements utilize only a single KIPA, the degree of vacuum squeezing that is achieved is limited by noise added by the HEMT amplifier.

\subsection{Frequency Dependence of Pump Transmission}

\begin{figure*}
    \centering
    \includegraphics{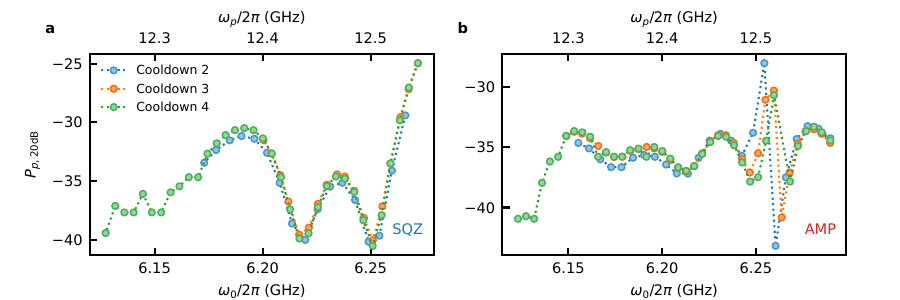}
    \caption{\textbf{Pump power needed for 20~dB of non-degenerate gain}. (\textbf{a}) $P_{p,\mathrm{20dB}}$ for SQZ measured as a function of the resonance frequency $\omega_0$. The pump power is plotted for three separate cooldowns. (\textbf{b}) An equivalent measurement for AMP.}
    \label{fig:pump_power_20_db}
\end{figure*}

In this section we infer the frequency-dependent transmission of the pump power to each device. We first tune each KIPA to a particular frequency $\omega_0$ using the DC current $\Idc$. Next, we measure the non-degenerate gain for that KIPA as a function of pump power $P_p$ with the pump frequency set to $\omega_p=2\omega_0$. We then determine the precise pump power required to achieve a non-degenerate gain of $20$~dB, which we define as $P_{p, \mathrm{20dB}}$. Finally, we plot the frequency-dependence of $P_{p, \mathrm{20dB}}$, which we measure across the operational frequency range of each KIPA (Fig.~\ref{fig:pump_power_20_db}). 

Both KIPAs show up to $15~$dB variations in $P_p(\omega_0)$ over small frequency ranges, which we attribute to reflections in the lines and device at the pump frequency. Despite this, Fig.~\ref{fig:pump_power_20_db} shows that for both devices $P_{p,\mathrm{20dB}}(\omega_0)$ is reproducible and stable across several thermal cycles of the dilution refrigerator. This demonstrates that the KIPAs have excellent stability and that the pump frequency dependence can be calibrated. 

\FloatBarrier

\subsection{Variation of Resonance Frequency and Quality Factors with Magnetic Field}

\begin{figure*}[t!]
\includegraphics[width=0.85\textwidth]{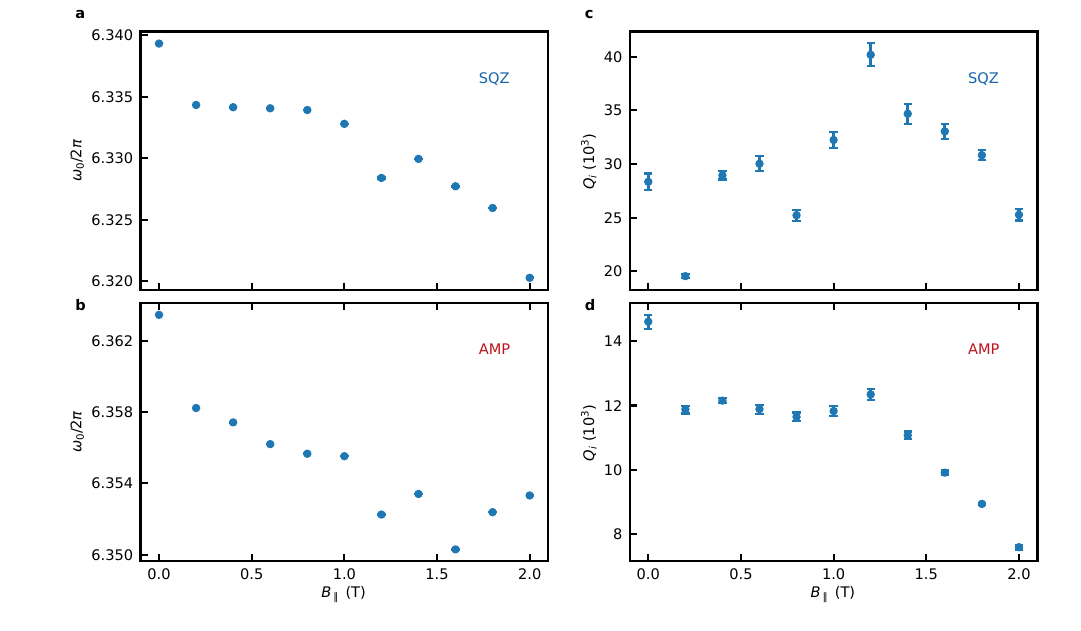}
\caption{\textbf{Field dependence of KIPAs} (\textbf{a,b}) Magnetic field $B_\parallel$ dependence of the resonance frequency $\omega_0/2\pi$ of SQZ (\textbf{a}) and AMP (\textbf{b}). A gradual decrease in $\omega_0$ is observed for both KIPAs with a 20~MHz decrease for the SQZ and 10~MHz decrease for the AMP at $B_\parallel=2$~T. (\textbf{c,d}) Internal quality factor $Q_i$ measured as a function of $B_\parallel$. We observe a decrease in $Q_i$ for both KIPAs, however, AMP is degraded to a lower $Q_i$ ($8\times 10^3$) compared to SQZ ($20\times 10^3$). }
\label{fig:q vs field}
\end{figure*}

As part of the experiment presented in Figs.~4a-c of the main text, we extract $\omega_0$ and $Q_i$ for SQZ and AMP at zero current ($\Idc=0$) as a function of the strength of the magnetic field applied parallel to SQZ $B_\parallel$ (Fig.~\ref{fig:q vs field}). The parameters are extracted from measurements of $S_{11}$. For all $B_\parallel$, measurements of $S_{11}$ were taken with a signal power of $-126$~dBm (based on a calibration of the fixed attenuation and line loss using the noise measurements in Section \ref{SI:noise_temp_meas}) referred to the input of the device. We estimate this to result in an average intracavity photon number below one based on the quality factors and applied signal power~\cite{Bruno2015}. Throughout the measurements, the coupling quality factors $Q_c$ of SQZ and AMP were found to remain stable. To keep the net magnetic field aligned in plane with SQZ, each time we increase the field of the primary magnet coil we adjust the field produced by an orthogonal coil which is nominally aligned to be out-of-plane with SQZ, to compensate for any misalignment. We use $\omega_0$ of SQZ as an indicator for the field alignment~\cite{Healey2008}, where $\omega_0$ is maximal for a field aligned along $B_\parallel$.

A gradual decrease in $\omega_0$ for both KIPAs is observed as $B_\parallel$ is increased (Fig.~\ref{fig:q vs field}a,b). This response was expected for SQZ because magnetic fields are known to increase kinetic inductance by decreasing the supercurrent density~\cite{Healey2008}. AMP is fixed to the MXC plate well outside of the magnet bore. Nevertheless, the stray magnetic field at its position is expected to be of order 10~mT for $B_\parallel=2~T$~\cite{bluefors}. While the stray field is significantly weaker in magnitude than $B_\parallel$, its alignment is predominantly out-of-plane to AMP and thereby results in a similar shift in $\omega_0$.

The extracted internal quality factors are also found to degrade gradually with $B_\parallel$. Interestingly, this effect is more pronounced for AMP than it is for SQZ. This indicates that for the experiments shown in Fig.~4a-c of the main text, the stray field impinging on AMP was likely a key limitation to the degree of squeezing achieved. This is further supported by the fact that we could not proceed beyond $B_\parallel>2~$T, because the gain of AMP could not be made to exceed 34~dB without the device turning normal.

\subsection{Variation of Resonance Frequency and Quality Factor with Temperature}

\begin{figure*}[t!]
\includegraphics[width=0.85\textwidth]{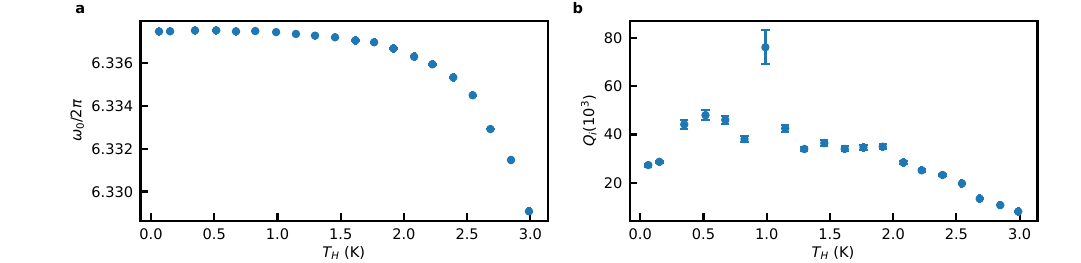}
\caption{\textbf{Temperature dependence of SQZ} (\textbf{a}) The resonance frequency $\omega_0$ declines monotoncially with $\Th$, as expected from the temperature dependence of the kinetic inductance. (\textbf{b}) Internal quality factor $Q_i$ as a function of temperature.}
\label{fig:q vs temperature}
\end{figure*}

In Fig.~\ref{fig:q vs temperature} we show $\omega_0$ and $Q_i$ for SQZ, measured as a function of temperature $T_H$ during the experiment presented in Figs.~4d-f of the main text. As in the previous section, these parameters are extracted from measurements of the scattering parameter $S_{11}$. The monotonic decline in $\omega_0$ with $T_H$ is consistent with the temperature dependence of the kinetic inductance~\cite{Annunziata2010}. Correspondingly, we modify $\Idc$ for each $T_H$ during the squeezing measurements in Figs.~4d-f to maintain a constant $\omega_0$ throughout the experiment. 

\subsection{Amplifier Added Noise}
\label{SI:noise_temp_meas}

\begin{figure*}
    \centering
    \includegraphics[width=\textwidth]{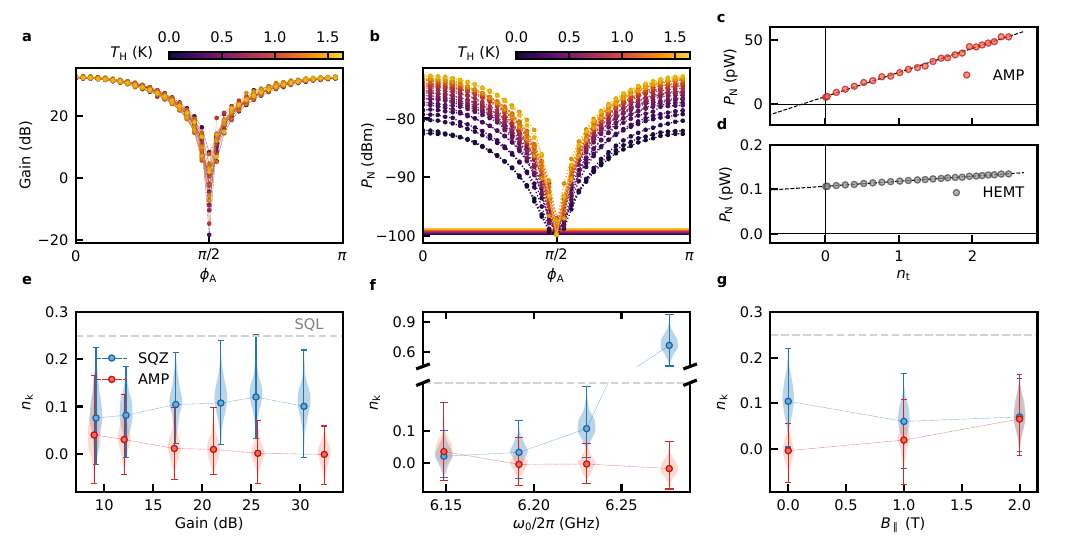}
    \caption{\textbf{Amplifier added noise characterization in anti-squeezing mode.} (\textbf{a}) The phase-dependent gain of AMP, measured for noise source temperatures $\Th$ in the range 60~mK to 1.6~K. Note that the gain is stable and independent of $\Th$. (\textbf{b}) The corresponding phase-dependent (de-)amplified noise. The output noise $\PN$ increases as the thermal noise source temperature is increased. The colored horizontal lines indicate the output power when AMP is deactivated. We use the data collected at $\phiamp=0$ for the determination of the added noise in anti-squeezing mode. (\textbf{c}) The noise power $\PN$ vs. number of thermal photons $\nth$ (for AMP). The noise added by the amplifier is calculated from the y-intercept and the slope. (\textbf{d}) The noise power vs. $\nth$ when both SQZ and AMP are deactivated. Using this measurement we determine the HEMT added noise as $n_\mathrm{H} = 6.9$ photons per quadrature. (\textbf{e}) Amplifier added noise vs. gain (at $\omega_0/2\pi=6.23$ GHz). The horizontal dashed line indicates the standard quantum limit (SQL) of 1/4 photon per quadrature. (\textbf{f}) Amplifier added noise vs. $\omega_0$ (at SQZ Gain = 18~dB and AMP Gain = 34~dB). (\textbf{g}) Amplifier added noise vs. parallel magnetic field $B_{\parallel}$ (at $\omega_0/2\pi=6.2301$~GHz, SQZ Gain = 18~dB and AMP Gain = 34~dB).}
    \label{fig:noise_temp}
\end{figure*}

As described in Section~\ref{SI:noise_temp_meas}, the procedure we use for characterizing the noise added by the KIPAs involves measuring the amplified noise power $\PN$ on a single quadrature as a function of the thermal noise source temperature $\Th$, or equivalently the number of thermal photons per quadrature of the input field $\nth$. This procedure is illustrated for AMP in Figs.~\ref{fig:noise_temp}a-d. At each $\Th$, we measure the phase-dependent gain of the KIPA using coherent states (to ensure its operation is stable, Fig.~\ref{fig:noise_temp}a) and the pump-phase-dependent noise power $\PN(\phiamp)$ (Fig.~\ref{fig:noise_temp}b).
The noise added by the KIPA, stated in terms of input-referred photons $n_k$, can then be found from the slope and intercept of the amplified noise power $\PN(\nth)$ (Fig.~\ref{fig:noise_temp}c), in accordance with the input-output model (Eqs.~\ref{eq:noise_temp_sqz}-\ref{eq:noise_temp_amp}). This also requires that we measure the noise power as a function of $\nth$ with both SQZ and AMP deactivated (Fig.~\ref{fig:noise_temp}d), from which we determine that the HEMT adds 6.9~photons of noise to each quadrature. For AMP, the ratio of the slope and y-intercept yields $\frac{1}{\eta_0\eta_1}\left[\frac{1}{4} + \na\right]$ (Eq.~\ref{eq:noise_temp_amp}). 
We assume the thermal occupation at the ports of the circulator $n_{\eta 0}=n_{\eta 1}=0$, which is justified because the circulator is thermalized to the mixing chamber of the dilution refrigerator, which is kept below 44~mK throughout the experiment, and all lines of the dilution refrigerator are heavily attenuated in the relevant frequency band.

We take into account uncertainties in the measured data points and the parameters $\eta_0$ and $\eta_1$ with a bootstrapping approach. The coordinates of each data point are sampled from a normal distribution around the mean, with a standard deviation given by the propagated measurement error in the $\nth$ and $\PN$ coordinates. The parameters $\eta_0$ and $\eta_1$ are sampled from a normal distribution with a $1\sigma$ confidence interval ranging from 0.85 to 1, based on the fits and uncertainties in Figs.~3d,e of the main text. The noise figure $n_k$ is then determined in $N=400$ samples by applying a weighted fit through the sampled coordinates, with weights inversely proportional to the data point variance. Taking the mean of the bootstrap samples then gives the estimate for $n_k$.

The amplifier added noise estimate $n_k$ and its probability distribution are plotted in Figs.~\ref{fig:noise_temp}e-g as a function of three parameters: (anti-squeezing) gain, frequency $\omega_0/2\pi$ and parallel magnetic field $B_{\parallel}$. At the optimal operational gains ($34$~dB and $18$~dB for AMP and SQZ, respectively) used in Fig.~3 of the main text, we estimate the single-quadrature noise figure in anti-squeezing mode to be $n_{\mathrm{A}}^{\mathrm{anti}}= 0.00 \pm 0.02$ and $n_{\mathrm{S}}^{\mathrm{anti}} = 0.11 \pm 0.02$. It must be noted here that the noise figure of interest for SQZ is in squeezing, instead of anti-squeezing mode. We determine the noise figure in squeezing mode to be $n_{\mathrm{S}}^{\mathrm{sq}} = 0.02 \pm 0.02$ by fitting the data in Fig.~4g of the main text with Eq.~\ref{eq:squeezing_thermal}. Such asymmetries in the added noise between squeezing and anti-squeezing mode have been previously documented, in both theory~\cite{Caves1982} and experiment~\cite{Zhong2013}.

\clearpage
\section{Supporting Squeezing Data}

\subsection{Phase alignment for the squeezing measurements}

\begin{figure*}[h]
    \centering
    \includegraphics{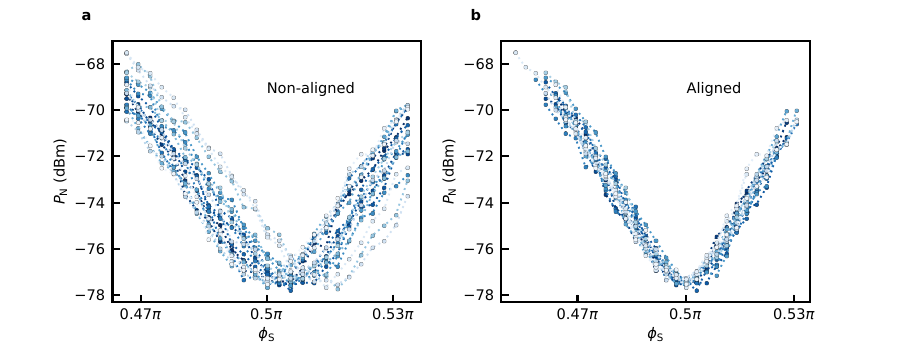}
    \caption{\textbf{Alignment procedure for the squeezing measurements, accounting for slow phase drifts.} (\textbf{a}) Raw measured data of the noise power $\PN$ for Fig.3c of the main text. The $\PN$ reaches approximately the same minimum for each measurement repetition, with a minor variation in the pump phase. (\textbf{b}) The raw data is aligned by fitting each repetition with a parabolic function and centring the minima on $\phisqz=\pi/2$.}
    \label{fig:alignment}
\end{figure*}

\noindent We account for slow phase drifts in the microwave pump sources by centering separate measurement traces, as displayed in Fig.~\ref{fig:alignment}.

\FloatBarrier

\FloatBarrier

\subsection{Estimation of insertion loss and added noise photons in the thermal squeezing configuration}

\begin{figure}[h]
    \centering
    \includegraphics{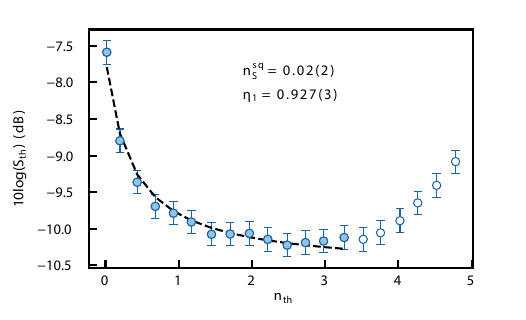}
    \caption{Fit of thermal squeezing vs. number of thermal noise photons $n_{\mathrm{th}}$ to the thermal squeezing model in Supplementary Eq.~\ref{eq:squeezing_thermal}. The fit result yields the amplifier added noise $n_{\mathrm{S}}^{\mathrm{sq}}$ and insertion loss $\eta_1$ for the setup displayed in Fig.~4d of the main text.}
    \label{fig:thermal_squeezing_fit}
\end{figure}

\clearpage
\subsection{Establishing the vacuum level in the thermal squeezing measurments}

\noindent Due to the increased heat load of the modified setup depicted in Fig.~4d of the main text, the minimum $T_\mathrm{H}$ we could reach was 80~mK, as opposed to 10~mK in the regular squeezing measurements. Therefore, we obtain the vacuum level by extrapolating $\Pa$ to $T_\mathrm{H}=0$~K. The extrapolated $\Pvac=79.9$~dBm is slightly lower than the first measured value at $T=80$~mK ($n_\mathrm{t}=0.017$, $\PN=-79.61$~dBm), and thus prevents overestimating the squeezing level in Fig.4f of the main text.

\begin{figure*}[h]
    \centering
    \includegraphics{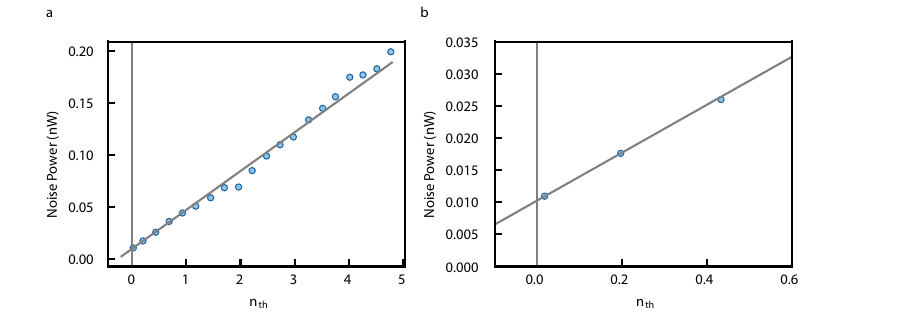}
    \caption{\textbf{Noise extrapolation to $n_{\mathrm{th}}$ for determining the vacuum level in the thermal squeezing measurements.} (\textbf{b}) Noise power vs. thermal noise photons per quadrature $n_{\mathrm{th}}$. In this measurement, both SQZ and AMP are turned off. We observe a linear increase in the noise power as a function of the number of thermal noise photons. (\textbf{b}) Zoom-in of panel \textbf{a}. We extrapolate a linear fit back to $n_{\mathrm{th}}=0$ to obtain the vacuum noise reference of 0.1025~nW, or $\Pvac=-79.9$~dBm.}
    \label{fig:extrapolation}
\end{figure*}

\FloatBarrier

\subsection{Frequency and Bandwidth Dependence of Squeezing}

\begin{figure*}[h]
    \centering
    \includegraphics{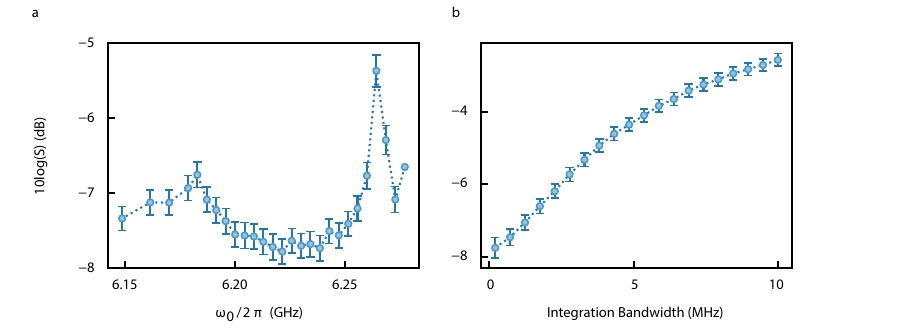}
    \caption{\textbf{Squeezing as a function of frequency and integration bandwidth.} (\textbf{a}) Squeezing as a function of the resonance frequency $\omega_0$ of the KIPAs. We demonstrate squeezing over a range of approximately 150~MHz. For each data point, the gain of SQZ is optimized to reach the maximum squeezing level. The gain of AMP is stabilized at 34~dB. We achieve maximum vacuum squeezing at $\omega_0/2\pi = 6.230$~GHz, corresponding to $I^{\mathrm{S}}_{\mathrm{DC}}=0.90$~mA and $I^{\mathrm{A}}_{\mathrm{DC}}=0.97$~mA. The frequency range is limited by the critical currents of the superconducting films. (\textbf{b}) Squeezing as a function of the integration bandwidth.}
    \label{fig:frequency_dependence}
\end{figure*}

\FloatBarrier

\subsection{Pump Crosstalk}


In the squeezing experiments we tune SQZ and AMP to a mutual resonance frequency $\omega_0$ using DC currents, and supply independent pump tones with a frequency $\omega_p=2\omega_0$. To investigate the pump crosstalk between SQZ and AMP, we configure both devices as if they were to be used in a squeezing measurement, and then deactivate one device by setting its current to $\Idc=0$. The two pump tones remain on with a mutual frequency of $\omega_p=2\omega_0$. We then measure the amplified noise power $\PN$ as a function of the pump phases $\phisqz$ and $\phiamp$.

To characterize this crosstalk, we use the configuration where we obtained the largest degree of squeezing (-7.8(2)~dB, Fig.~3 of the main text). This occurs at $\omega_0/2\pi=6.23~$GHz, with pump powers (maximum degenerate gains) of $P_{\mathrm{p, S}}=-37$~dBm (18~dB) and $P_{\mathrm{p, A}}=-34$~dBm (34~dB) for SQZ and AMP, respectively.

\begin{figure*}[h]
    \centering
    \includegraphics[width=0.65\textwidth]{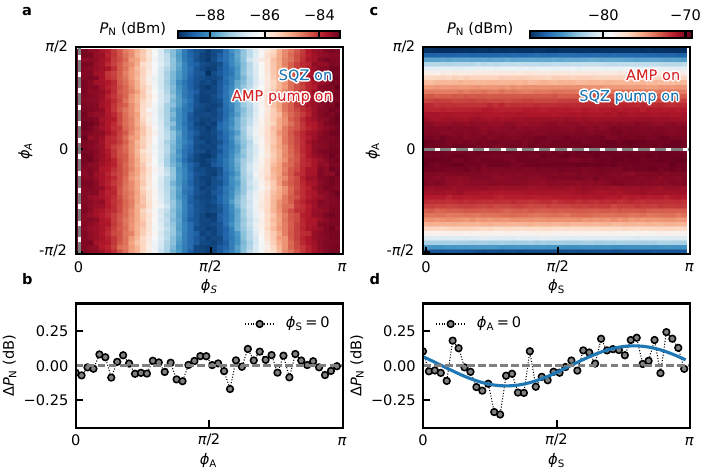}
    \caption{\textbf{Measurement of pump cross talk between the KIPAs.} (\textbf{a}) Measurement of the influence of the AMP pump on the noise amplification of SQZ. We plot the noise power $\PN$ as a function of the pump phases $\phiamp$ and $\phisqz$ with SQZ activated and AMP tuned out of resonance ($\Idc$ = 0). The AMP pump power is set at $P_{\mathrm{p, A}} 
    = -34$~dBm, the same setting used for optimal squeezing in Fig. 3 of the main text.
    (\textbf{b}) A linecut at $\phisqz=0$ of $\PN$, plotted as a variation about its mean value. We observe no dependence of $\PN$ with $\phiamp$.
    (\textbf{c}) Measurement of the influence of the SQZ pump on the noise amplification of AMP. The measurement is the same as in panel (\textbf{a}) with the roles of SQZ and AMP reversed.
    (\textbf{d}) A linecut of $\PN$ at $\phiamp=0$, plotted as a variation about its mean, reveals a small sinusoidal dependence on $\phisqz$ with an amplitude of 0.15~dB.}
    \label{fig:crosstalk_pump}
\end{figure*}

In Fig.~\ref{fig:crosstalk_pump}a we measure $\PN(\phisqz,\phiamp)$ with SQZ activated and AMP deactivated. We observe no dependence on $\PN$ with $\phiamp$, as exemplified by Fig.~\ref{fig:crosstalk_pump}b, which shows the line-cut $\PN(\phiamp)|_{\phisqz=0}$ (for clarity, we plot it as a variation about its mean value). We thus conclude that the pump of AMP has no measurable effect on SQZ. In Fig.~\ref{fig:crosstalk_pump}c we measure $\PN(\phisqz,\phiamp)$ in the reverse configuration, with SQZ deactivated and AMP activated. The line-cut $\PN(\phisqz)|_{\phiamp=0}$ reveals a weak dependence on $\phisqz$. We fit the line-cut with a sinusoid and extract an amplitude of 0.15~dB. Whilst this phase-dependent cross talk does not affect our evaluation of the vacuum noise (which is averaged over all $\phisqz$), it may impact our measurement of the squeezed noise level by $\pm 0.15$~dB. We take this into account as an additional systematic uncertainty in our squeezing estimates.

\FloatBarrier
